\documentclass{emulateapj}
\usepackage{apjfonts}
\usepackage{graphics}

\slugcomment{}

\shorttitle{HDE 245059: A binary revealed by Chandra \& Keck}
\shortauthors{Baldovin Saavedra et al.}

\begin{document}


\title{HDE 245059: a weak-lined T Tauri binary revealed by Chandra \& Keck }


\author{C. Baldovin Saavedra\altaffilmark{1,2}}

\author{ M. Audard\altaffilmark{1,2}}

\author{G. Duch\^{e}ne\altaffilmark{3,4}}

\author{M. G\"{u}del\altaffilmark{5}}

\author{S.L. Skinner\altaffilmark{6}}

\author{F. B. S. Paerels\altaffilmark{7}}

\author{A. Ghez\altaffilmark{8}}
\author{C. McCabe\altaffilmark{9}}

\altaffiltext{1}{ISDC, Universit\'{e} de Gen\`{e}ve, Chemin d'Ecogia 16, 1290 Versoix, Switzerland}
\altaffiltext{2}{Observatoire de Gen\`{e}ve, Universit\'{e} de Gen\`{e}ve, Chemin des Maillettes 51, 1290 Versoix, Switzerland}
\altaffiltext{3}{Laboratoire d'Astrophysique de l'Observatoire de Grenoble, 414 Rue de la Piscine, Domaine Universitaire, BP 53, 38041, Grenoble Cedex 09, France}
\altaffiltext{4}{Astronomy Department, UC Berkeley, 601 Campbell Hall, Berkeley, CA 94720-3411, USA}
\altaffiltext{5}{Institute of Astronomy, ETH Z\"{u}rich, CH-8092 Z\"{u}rich, Switzerland}
\altaffiltext{6}{CASA, University of Colorado, 389 UCB, Boulder, CO 80309-0389, USA }
\altaffiltext{7}{Columbia Astrophysics Laboratory, 550 West 120th Street, New York, NY 10027, USA}
\altaffiltext{8}{UCLA, 430 Portola Plaza, Box 951547, Los Angeles, CA 90095-1547, USA}
\altaffiltext{9}{IPAC, California Institute of Technology, 1200 East California Boulevard Pasadena, CA 91125, USA}


\begin{abstract}

We present the \emph{Chandra} High Energy Transmission Grating Spectrometer (HETGS) and \emph{Keck} observations of HDE 245059, 
a young weak-lined T Tauri star (WTTS), member of the pre-main sequence group in the $\lambda$~Orionis Cluster. 
Our high spatial resolution, near-infrared observations with Keck reveal that HDE 245059 is in fact a binary separated by 0\farcs87, probably 
composed of two WTTS based on their color indices. 
Based on this new information we have obtained an estimate of the masses of the binary components; 
$\approx 3\,M_\odot$ and $\approx2.5\,M_\odot$ for the north and south components, respectively.
We have also estimated the age of the system to be $\approx2-3$\,Myr. 
We detect both components of the binary in the zeroth order \emph{Chandra} image and in the grating spectra. 
The lightcurves show X-ray variability of both sources and in particular a flaring event in the weaker southern component. 
The spectra of both stars show similar features: a combination of cool and hot plasma as demonstrated by several iron lines 
from Fe XVII to Fe XXV and a strong bremsstrahlung continuum at short wavelengths. 
We have fitted the combined grating and zeroth order spectrum (considering the contribution of both stars) 
in XSPEC. 
The coronal abundances and emission measure distribution for the binary have been obtained using different methods, including a continuous emission measure distribution and a multi-T approximation. 
In all cases we have found that the emission is dominated by a plasma between $\sim8$ and $\sim15$~MK a soft component at $\sim 4$ MK and a hard component at $\sim 50$ MK are also detected.
The value of the hydrogen column density was low, N$_{\rm H} \sim 8 \times 10^{19}$~cm$^{-2}$, likely due to the clearing of the inner region of the 
$\lambda$ Orionis cloud, where HDE~245059 is located. 
The abundance pattern shows an inverse First Ionization Potential (FIP) effect for all elements from O to Fe, the only exception being Ca.
To obtain the properties of the binary components, a 3-$T$ model was fitted to the individual zeroth order spectra using the abundances derived for the binary.
We have also obtained several lines fluxes from the grating spectra. 
The fits to the triplets show no evidence of high densities.
In conclusion, the X-ray properties of the weak-lined T Tau binary HDE 245059 are similar to those generally observed in other weak-lined T Tau stars. 
Although its accretion history may have been affected by the clearing of the interstellar material around $\lambda$~Ori, its coronal properties appears
not to have been strongly modified.

\end{abstract}

%
%
%
%
%
%
 \section{Introduction}\label{intro}

T Tauri stars (TTS) are optically revealed pre-main sequence stars (PMS) of low mass ($M_{*} \sim 0.2 - 3~ M_\odot$), whose interiors are fully convective and powered 
principally by gravitational contraction rather than by nuclear reactions. 
Low-mass PMS stars are classified optically in two types: classical T Tauri stars (CTTS) and weak-lined T Tauri stars (WTTS).
The CTTS present strong H$\alpha$ emission lines, a signature of accretion, and infrared excess, revealing a dusty circumstellar disk. 
On the other hand, WTTS present weaker H$\alpha$ emission lines and little or no infrared excess, which is generally interpreted as a sign that active accretion 
has significantly decreased and the disk has become optically thin.  
PMS stars are magnetically active and rotate slowly during the early stages of their evolution and spin up as they contract to the main sequence \citep{stauff86}.
Recent evidence suggests that there is also a population of low-mass PMS stars ($0.3 \leq M/M_\odot \leq 1$) that seem to contract towards the zero-age main
sequence (ZAMS) at constant angular velocity during the first 3-5~Myr after they begin their evolution down the convective tracks \citep{rebull04}, 
indicating there should be a mechanism extracting angular momentum in these PMS objects. 
In CTTS the presence of disk and magnetic fields is believed to keep the rotation rate low, while the short rotational period in WTTS may be due to the dispersal of the circumstellar disk.
But observational evidence has been sometimes confusing and the relation between accretion and angular momentum loss cannot be assured \citep{bouv07}. 

Both type of stars are strong X-rays emitters that were first detected with the \emph{Einstein} satellite, e.g. \citet{feig81} and 
subsequent observations with \emph{ROSAT} (e.g. Feigelson et al. 1993, Neuh\"auser et al. 1995). 
Spectral measurements of X-rays from YSOs reveal emission from continuum and lines from an optically thin plasma. 
Most of the TTS show variability in X-rays with time scales of minutes to days and can present strong flares.

X-ray spectroscopy of CTTS has shown some evidence that their X-ray emission could be due, at least in part, to accretion shocks on the stellar 
photosphere \citep{kastner02a, stelzer04, schmitt05, gunther06,argiroffi07}, although some CTTS do not support this mechanism. 
Their X-ray emission is believed to be coronal \citep{audard05,gudel07,smith05}. 
A X-ray soft excess in CTTS is, however, reported compared to WTTS \citep{telleschi07a}.
X-rays in WTTS are thought to originate only from magnetic activity similar to main-sequence magnetically active stars, albeit at much higher levels. 

The low level of accretion in WTTS makes them the best-suited objects to study the magnetic activity in PMS stars, and the impact of stellar 
flares and X-rays onto their optically thin disks and their planets \citep{lammer03,lammer06,smith07}.

WTTS show strong X-ray and non-thermal radio emissions \citep{skinner93,feig99}.
Compared to zero-age main sequence stars, their X-ray luminosities are very high ($10^{28.5} - 10^{31}$~erg~s$^{-1}$), but they typically show 
$L_\mathrm{X}/L_\mathrm{bol} \approx 10^{-4}$. WTTS display very high coronal temperatures \citep{skinner97, tsuboi98}, X-ray flares with temperatures of a few tens of MK are frequently seen, e.g., \citet{tsuboi98,stelzer00}.  
Thanks to the \emph{Chandra X-ray Observatory} (CXO) and the \emph{X-ray Multi-Mirror Mission} (XMM~-~Newton), observations of PMS stars 
such as the WTTS HDE 245059 can provide a better understanding of magnetic activity and its evolution, 
allowing to resolve individual spectral lines and to get improved diagnostics of the plasma properties. 

CTTS and WTTS show similar ages and are found in the same area of the Hertzsprung-Russell diagram; why are they different then? 
The environment might be an interesting cause of the different properties. 
For example, a supernova explosion can clear away material, leading to the loss of the disks in some young stars (Dolan \& Mathieu 1999, 2001, 2002).
The presence of a massive star can also disperse the young circumstellar disks by photoevaporation due to UV radiation.
On the other hand, the above cannot hold for low-mass star forming regions, such as Taurus. However,
understanding the effect of environment onto young stars could give clues about the mechanism leading to the differentiation between CTTS and WTTS.

The main goals that we address in this work are: 
\emph{i)} to obtain the plasma properties; temperatures, abundances, X-ray luminosity of the WTTS HDE~245059 located in the $\lambda$ Ori star forming region, 
\emph{ii)} to obtain fluxes from individual lines, 
\emph{iii)} to derive age and masses of the system based on NIRSPEC data.

%
%
%
%
%
 \section{The HDE 245059 system}\label{hde245059}

The young binary HDE~245059 is a member of the $\lambda$ Ori cluster, a star forming region located at $d=400 \pm 40$ pc \citep{murdin77}. 
The $\lambda $ Ori cluster or Collinder 69 includes the O star $\lambda ^1$ Orionis, spectral type O8 III \citep{barrado07}. 
The star is near the center of the region which is believed to be the remnant of a supernova that exploded about 1 - 2 Myr ago. 
This explosion might have cleared out the region, leaving a ring of molecular gas surrounding $\lambda ^1$ Orionis \citep{dolan99,dolan01,dolan02}.
Star formation is, however, still active at the edges of the molecular ring.

The $\lambda $ Ori region is also populated by faint PMS stars and has been studied at different wavelengths; radio, infrared, visible,
UV \citep{maddalena87,dolan99, barrado07}. 
\citet{maddalena87} argued that the observed CO emission from transition $J=1\rightarrow 0$ together with UV, optical, IR, and 21 cm data of the surrounding molecular clouds  
indicates that the ring centered on $\lambda ^1$ Orionis is actually the remnant of a preexisting cloud. 
Dolan \& Mathieu (1999, 201, 2002) have studied the region and attempted to deduce its history.
According to the authors, about 6~Myr ago the star formation process started in the most massive clouds of the region giving birth to several OB stars.
This process increased gradually until approximately 1 Myr ago when a supernova exploded disrupting the central region and decreasing the star formation rate, yet not stopping it completely.   

Our target ($\lambda $ Ori X-1) was discovered in the X-rays with the \emph{Einstein} Observatory 
during observations of the region centered at $\lambda ^1$ Orionis. 
The photometric and astrometric study of the region shows evidence that our X-ray source is in fact member of the association. 
HDE~245059 shows a strong absorption in the Li (6707~\AA) line, an indicator of youth \citep{alcala00}. 
\citet{fernandez95} has measured the equivalent width of the H$\alpha$ emission line for HDE~245059, W(H$\alpha)=0.37$ \AA, indicating that the star is a WTTS. 
The star has a spectral type K1, its X-ray luminosity is high, $\log L_{\rm X} = 31.7$ ergs s$^{-1}$ \citep{stone85}. 
Based on comparison with evolutionary model isochrones, \citet{stone85} estimated a mass and age  of  $M= 2-3 ~M_{\odot}$ and  $t\sim 1-4$ Myr. 
More recent isochrones, e.g those by \citet{siess00} might, however, change the above estimates (see below, Section \ref{env-bin}).
\citet{skinner91} obtained an upper limit estimate for the disk mass around HDE~245059 of $0.32~M_\odot$ based on the 1.1~mm continuum emission.
\citet{padgett96} obtained the effective temperature of HDE~245059, $T_{\rm{eff}}=5410 \pm 110$~K, and an estimate of the photospheric iron abundance, [Fe/H]$=-0.07 \pm 0.13$. 
Our target is a fast rotator ($v$~sin~$i~\sim 25$~km~s$^{-1}$), which in general is a sign of stellar youth or binaries of spectral types G and K \citep{fekel97}.
The studies of \citet{fekel97} in the optical, and \citet{alcala00} in the X-ray did not find evidence of radial velocity variability, an indication that HDE ~245059 is not a spectroscopic binary.  
However, our high spatial resolution near-infrared observations with \emph{Keck} indicate that it is, in fact, a spatially separated binary (see Section \ref{ir-obs}).

%
%
%
%
%
 \section{Observations}\label{obs}
\subsection{Near-infrared}\label{ir-obs}

On December 13, 2003, we observed HDE 245059 with NIRSPEC \citep{mclean00} installed behind the adaptive optics system 
\citep{wizi00} on the 10m Keck II telescope. 
The imaging camera within NIRSPEC offers a pixel scale of 0\farcs0168$\pm$0\farcs0001 and an
absolute orientation of 1\fdg1$\pm$0\fdg8 as determined from observations of calibration binaries and of a reference field in the Orion Trapezium. 
HDE 245059 was used as the adaptive optics guide star with the wavefront sensor running at a frequency of 488Hz, resulting in good correction (Strehl ratios of 43
and 19\% at 2.2 and 1.6 $\mu$m, respectively). 
We acquired images of the system with the broadband $K$ ($\lambda_0=2.20~\mu$m, $\Delta \lambda=0.39~\mu$m) and NIRSPEC-5
($\lambda_0=1.61~\mu$m,  $\Delta \lambda=0.40~\mu$m, a close analog to the usual $H$ filter) filters as
well as with the narrow band Br$\gamma$ filter
($\lambda_0=2.165~\mu$m,  $\Delta \lambda=0.02~\mu$m). 
In each filter, a series of short exposures were coadded at four successive positions on the detector. 
The total integration times were 40s, 4s and 20s with the $K$, $H$ and Br$\gamma$ filters, respectively.
For each filter, the four independent images were medianed to create a
sky that was subtracted from each image. The images were then
corrected for flat-field effects and cosmetically cleaned. The
images were then realigned and averaged to produce the final
images. Relative astrometry and photometry was obtained using the
DAOPHOT package. The astrometric results from all three filters were finally
averaged to reduce random uncertainties.

\begin{figure}
\begin{center}
\includegraphics[angle=180,scale=0.3]{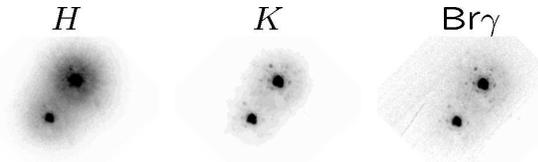}
\end{center}
\caption{Near-infrared Keck adaptive optics images of HDE 245059 in $H$, $K$, and Br$\gamma$ (2.165 $\mu$m) bands. The image is in square root scale, North is toward the top and East is toward the left.}
\label{fig:ir-image}
\end{figure}

The near-infrared images resolved HDE~245059 into a binary.
Figure \ref{fig:ir-image} shows the near-infrared Keck adaptive optics images of HDE~245059 with the three filters.
The northern component is brighter in all images.
The projected separation and uncertainty is $0\farcs 866\pm 0\farcs 005$ and the position angle, measured East from North
is 150\fdg0$\pm$1\fdg0.  
The final astrometric uncertainties are dominated by the absolute calibration uncertainties
of the detector. We measured magnitude differences of 0.98 mag, 0.88
mag and 0.86 mag with the $H$, $K$, and Br$\gamma$ filters respectively
(with typical uncertainties of 0.03 mag).
These differences suggest that both components have similar broadband colors. 

\subsection{X-rays}\label{xr-obs}

X-ray observations were performed using the \textit{High-Energy Transmission Gratings} (HETG) in combination with the \emph{Advanced CCD Imaging Spectrometer} (ACIS-S), 
on board of the \emph{Chandra X-Ray Observatory} (CXO). We obtained a total exposure time of 93 ks scheduled in three epochs: December 30, 2005, January 7, 2006 and January 13, 
2006 (Observation identification numbers 6241, 7253, and 5420, respectively). 
The short time interval between the observations allowed similar roll angles: 308\fdg5 for the first two and 294\fdg8 for the last one. 

Both components of the binary were detected in the zeroth order image. 
The binary orientation was close to the dispersion direction of the \emph{Chandra} MEG arm. 
Choosing the origin of the wavelength system between the two stars we were able to separate them in the grating spectra despite the small separation.

\section{Chandra Data Reduction and Analysis}\label{chandra}

The data of the HETGS observations were reduced from level 1 event files with the \emph{Chandra} Interaction Analysis of Observations software, CIAO 3.4, 
using the calibration database, CALDB 3.3.0.1, and following standard procedures to obtain a type 2 event file. 
We removed streak events which affect significantly ACIS-S4 (CCD ID=8). 

\subsection{Zeroth Order Images}
\label{zerothimage}

For the zeroth order images we used the subpixel event repositionary (SER) algorithm in order to improve spatial resolution \citep{tsunemi01,li03,li04}. 
When an X-ray photon hits the detector, the charge cloud created can either be spread into neighbouring pixels (called split pixel event)
or remains in a single pixel (called single pixel event).
Positions are determined with higher accuracy in the case of a split pixel event, however these events represent only a small fraction of the total events.
The SER technique uses both single pixel events and 2 pixel split events to increase the statistics.
When applied to the \emph{ACIS} observations the SER technique reduces the uncertainties in the determination of the photon impact position improving the spatial
resolution. The improvement in FWHM is typically between 40 to 70 \%. 

\begin{figure}
\begin{center}
\includegraphics[angle=0,scale=0.1]{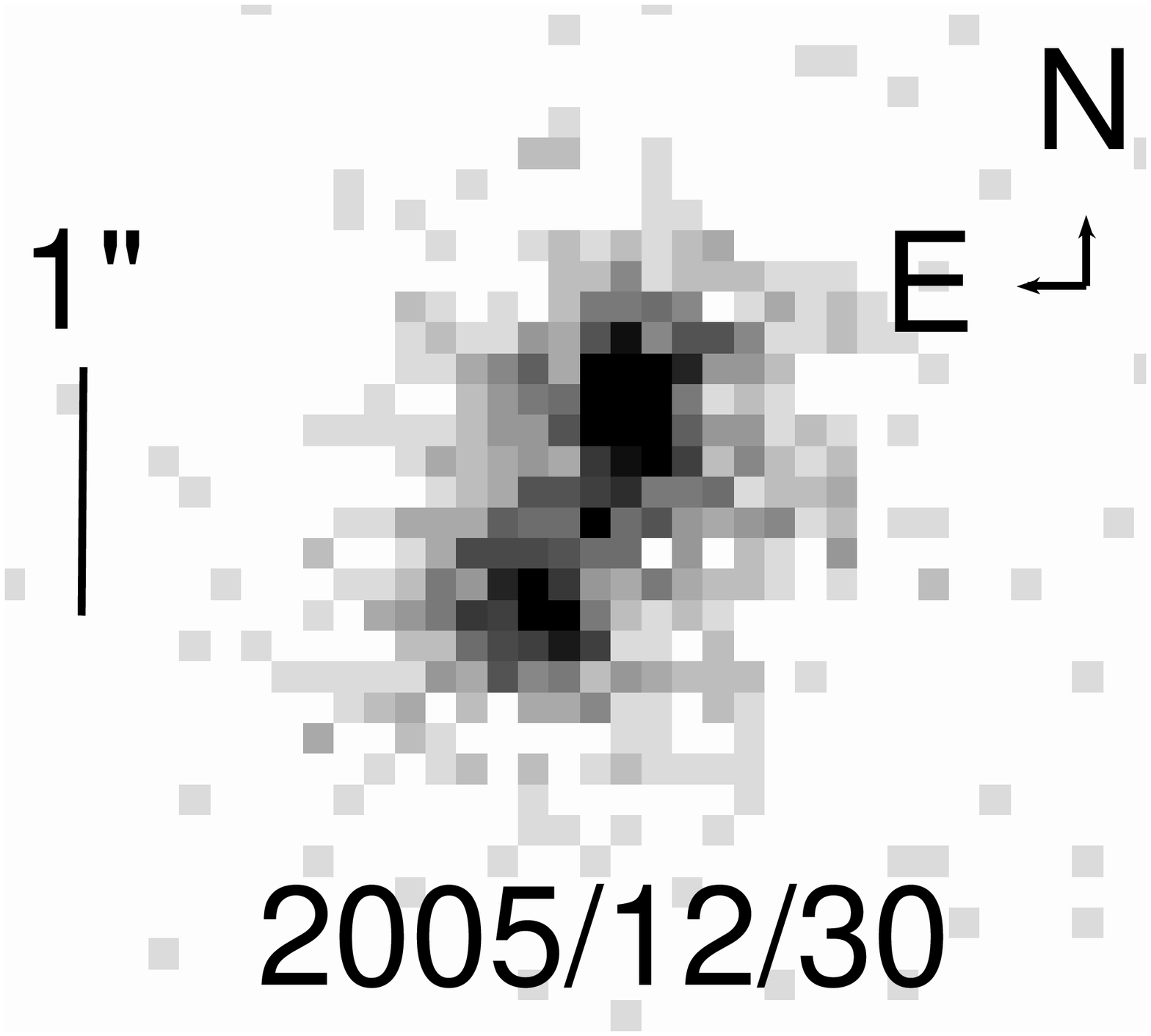}
\includegraphics[angle=0,scale=0.1]{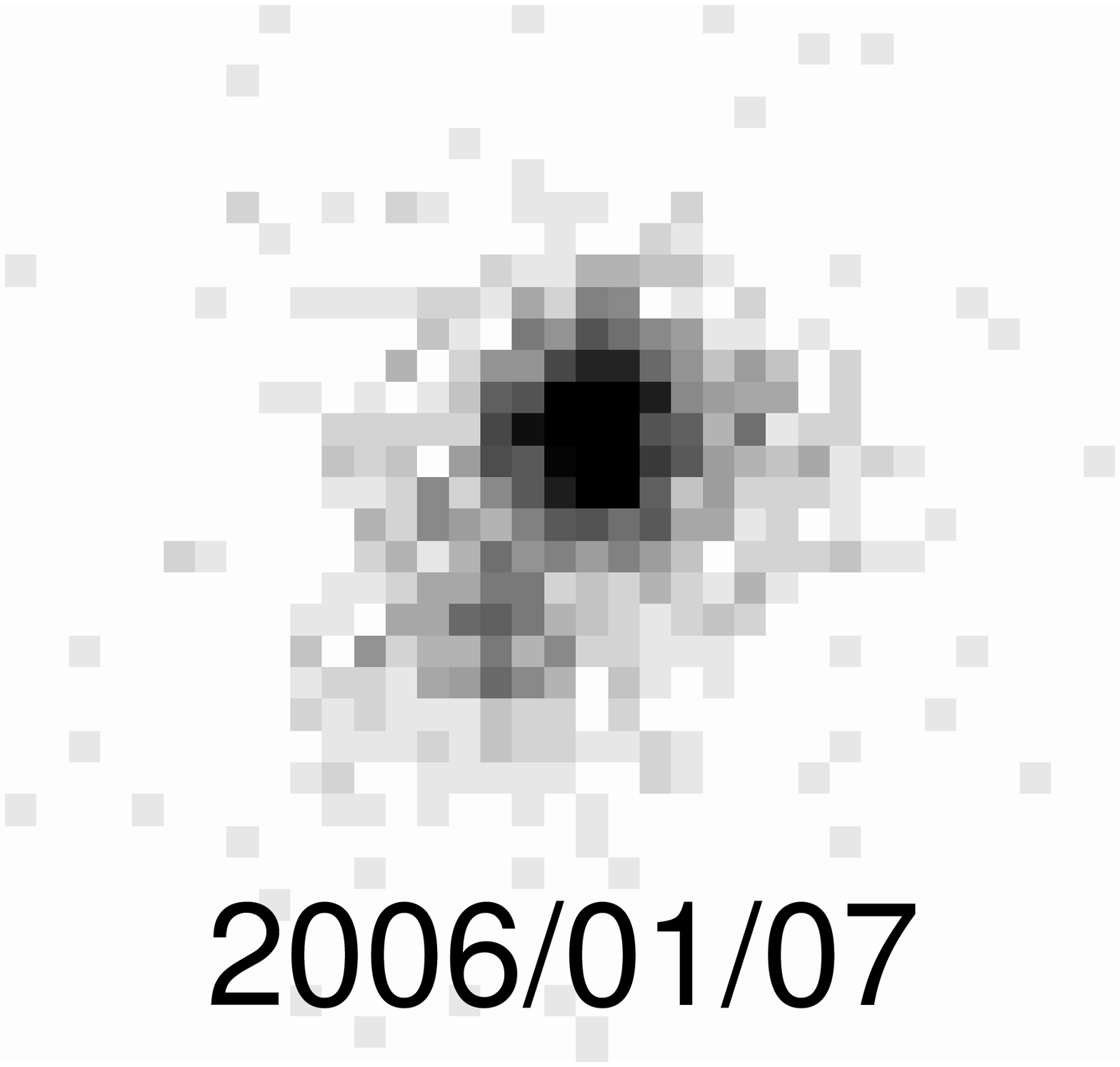} 
\includegraphics[angle=0, scale=0.1]{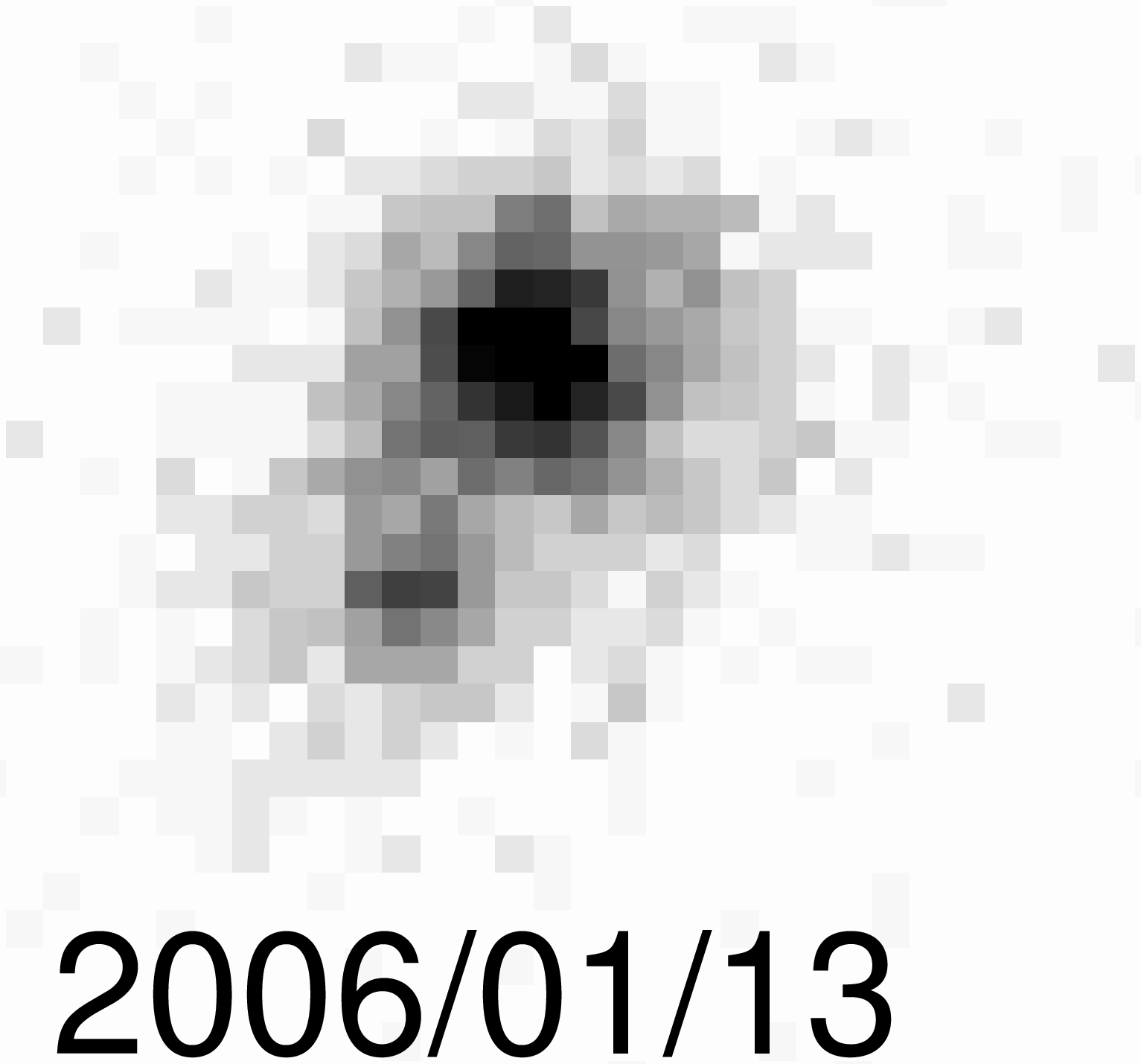}
\includegraphics[angle=0,scale=0.1]{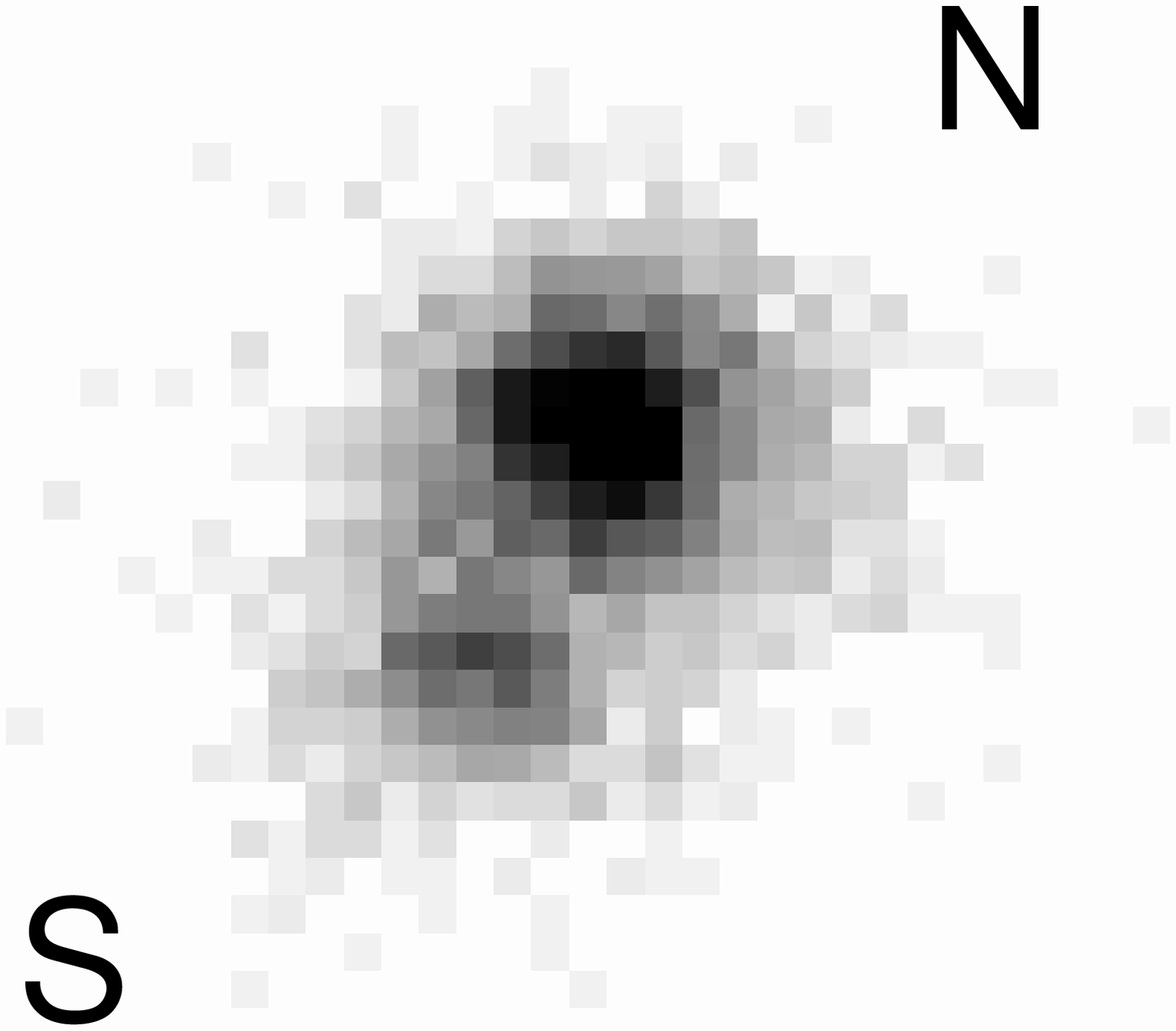}
\end{center}
\caption{Zeroth order X-ray image of HDE 245059 in square root scale for the three observation epochs. 
A flare is visible in the south component during the first epoch. At the far right, the average image obtained by summing the threee observations. 
}\label{fig:zerothimage}
\end{figure}

The zeroth order images of HDE~245059 for each observation epoch are displayed in Figure~\ref{fig:zerothimage}. 
A flare from the south component is visible in the first observation epoch (first panel). 
The summed zeroth order image over all the observation runs, Figure~\ref{fig:zerothimage} last panel, clearly shows the two components, the north component being on average the
brighter one. 

\subsection{Light Curves}
\label{lightc}

\begin{figure}
\begin{center}
\includegraphics[angle=0,scale=0.35]{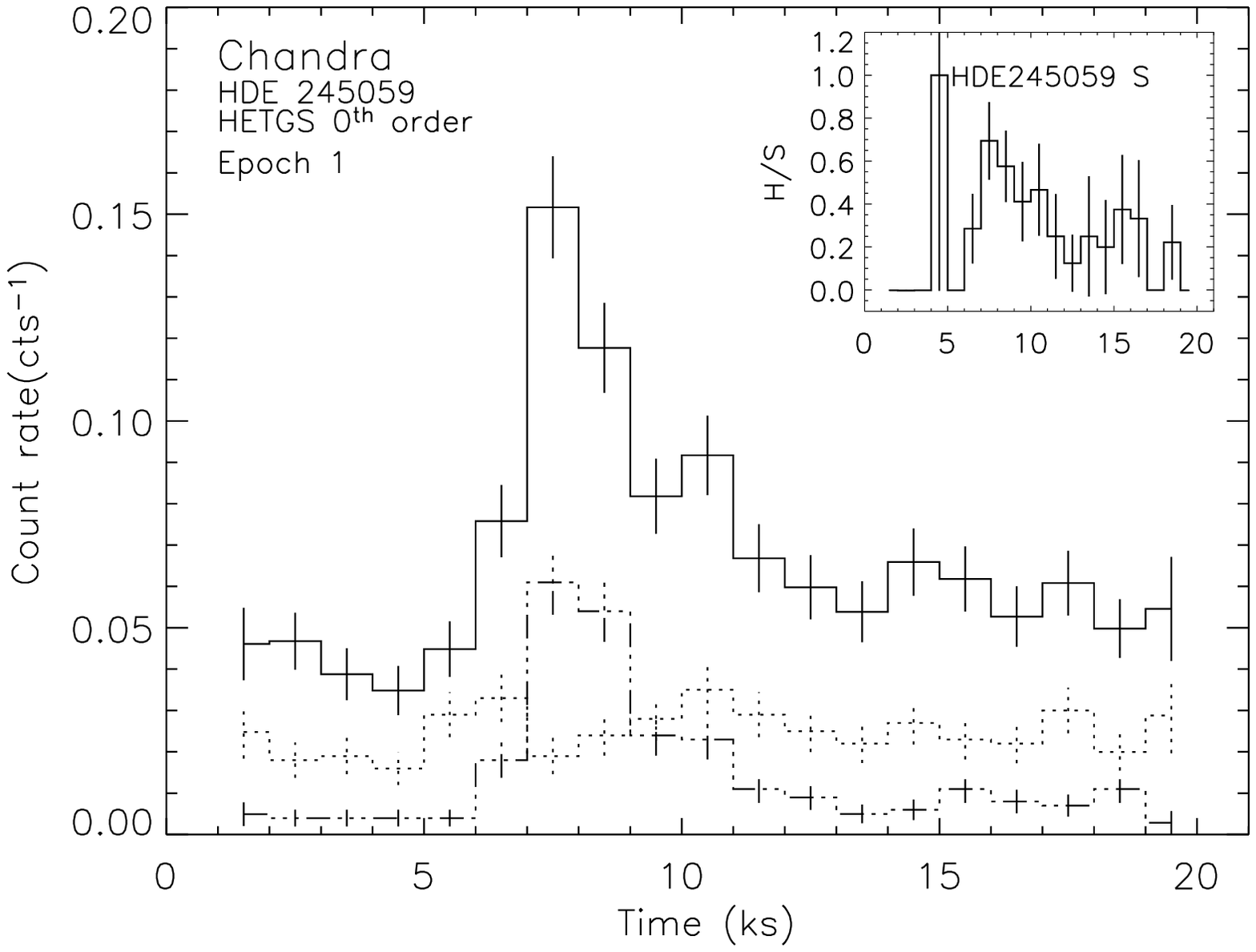}
\includegraphics[angle=0, scale=0.35]{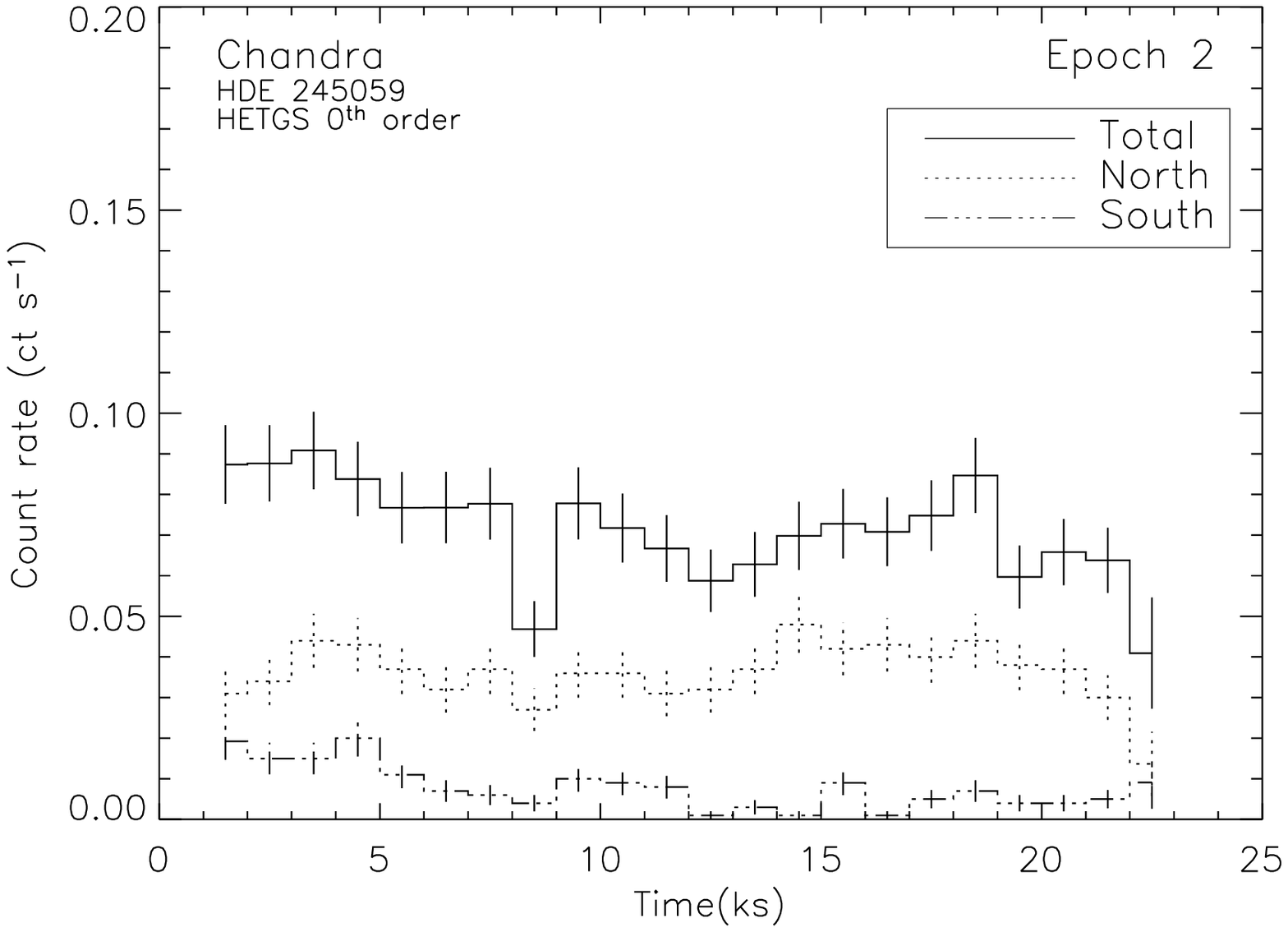}
\includegraphics[angle=0, scale=0.35]{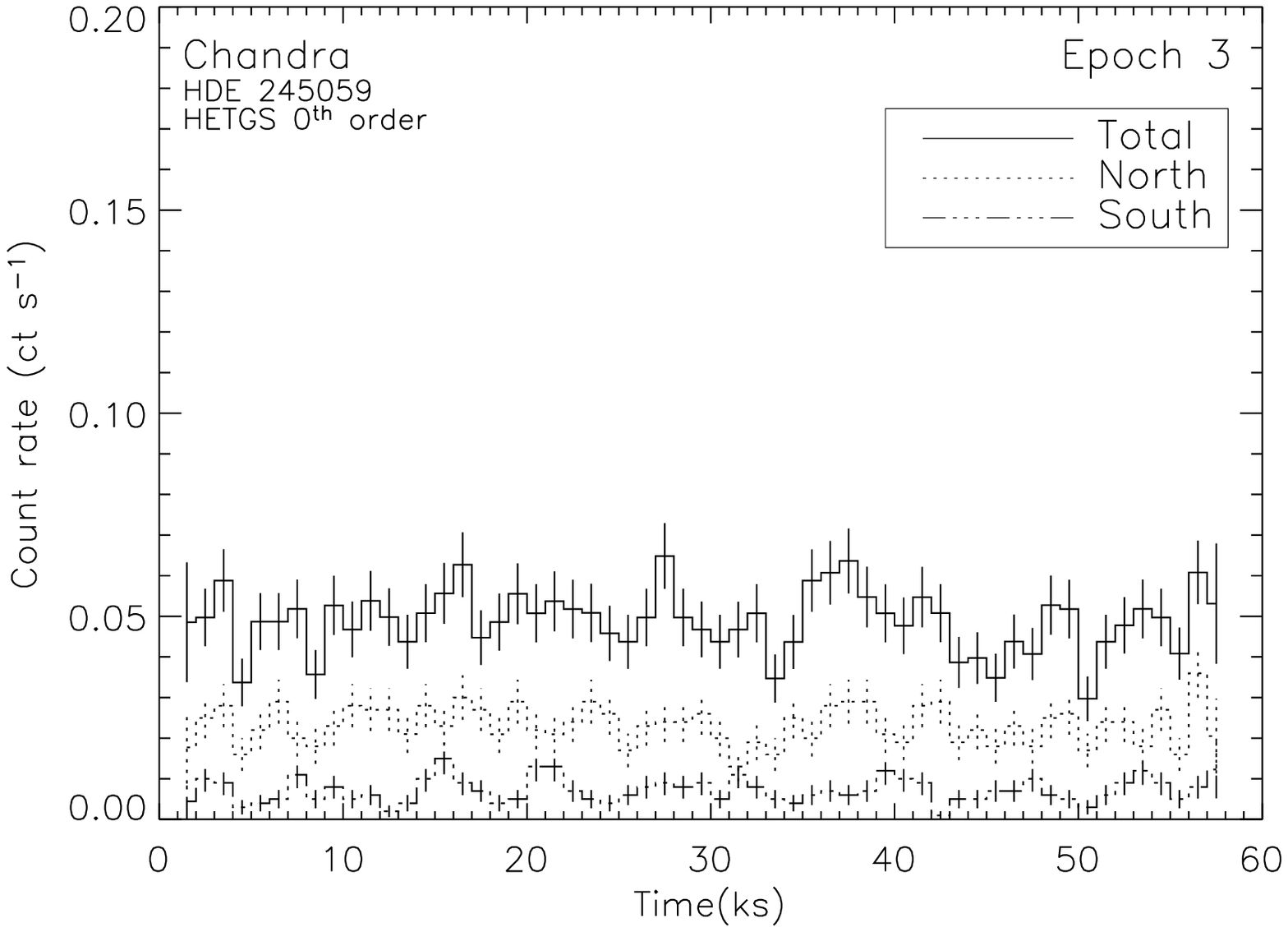}
\end{center}
\caption{Zeroth order lightcurve of HDE 245059 for the three observation epochs, binned by $10^{3}$ s. 
A flare is visible in the south component during the first epoch. The combined light curve of the binary is shown by the solid line, 
by the dotted line the contribution from the northern component and by the dashed-dotted line the contribution from the southern component.
For the first epoch lightcurve we have also included a plot of the hardness ratio for the southern star.}
\label{fig:lightczeroth}
\end{figure}

Figure~\ref{fig:lightczeroth} shows the light curves of the zeroth order data for the 3 observation epochs. 
The light curves were extracted from the data treated with the SER algorithm using a 2\farcs1 radius for the binary 
and 0\farcs4 radius for each component, thus, the sum of the lightcurves of both components do not match the binary lightcurve.
During the first observation epoch, a flare is detected from the south star, starting approximately 6 ks after the beginning of the 
observation. 
The flare lasted for nearly 6~ks, peaking 1.5 ks after its start with 0.06~cts~s$^{-1}$, roughly 5 times higher than the southern star 
average count rate for the first observation epoch.

We have also included the hardness ratio for the first observation epoch as an inserted figure in the lightcurve plot 
(Figure~\ref{fig:lightczeroth} top panel). 
 The hardness ratio was calculated using the expression H/S, where the soft band (S) was taken in the range from 0.3 to 2 keV 
 and the hard band (H) was taken in the range from 2 to 10 keV.

\subsection{Spectra}\label{spectra}
\subsubsection{Binary Components}\label{zcomponents}

Thanks to the spatial resolution of \emph{Chandra} we have separated the components of the HDE~245059 binary in the zeroth order spectrum.
The spectrum was extracted after merging the event files from the three observation epochs following standard CIAO 
procedures, using a circular region of radius 0\farcs4 for each star. 
The background was extracted from an annular region with radii of 4\farcs6 and 21\farcs \,
We have considered events within the range $0.2 - 7$~keV, where most of the signal is concentrated.
We have obtained the spectra for the quiescent state for each component.

\begin{figure}
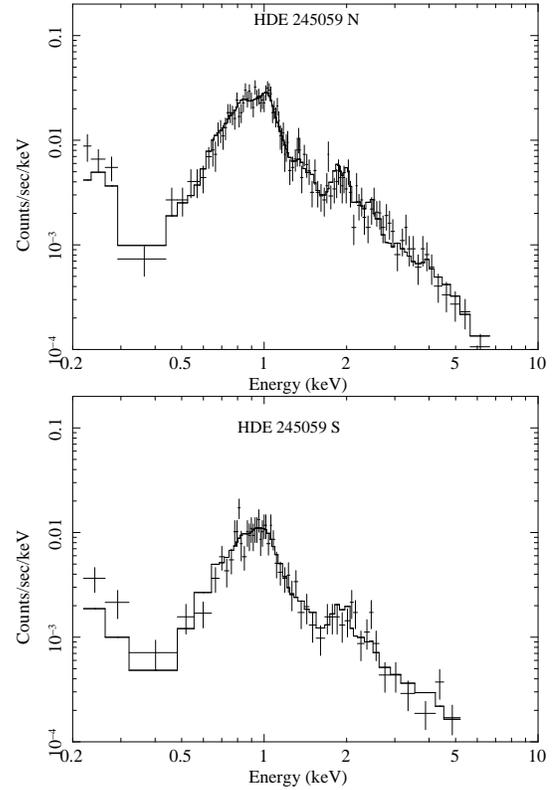

\begin{center}
\includegraphics[angle=270,scale=0.29]{HDE245059N-z.ps}
\includegraphics[angle=270,scale=0.29]{HDE245059S-q.ps}
\end{center}
\caption{Zeroth order average spectrum over the three observation epochs (in energy) for each component of the binary HDE 245059 (north and south).
We have overlaid the best-fit model (see Table \ref{table:ind_fits}) as a solid histogram.
}
\label{fig:zero_separate_spect}
\end{figure}

In Figure~\ref{fig:zero_separate_spect} we have plotted the spectra and overlaid the best-fit model for each star.
The northern star spectrum shows a higher signal with respect to the southern companion, which is consistent with the results from the zeroth order image.

We have fitted the individual spectra of the binary components obtained from the zeroth order data (see Table \ref{table:ind_fits}) using XSPEC version 11 \citep{arnaud96}, and a discrete emission measure distribution (EMD).
The hydrogen column density and abundances were fixed to the zeroth order plus grating best-fit values (discussed in section \ref{binaryspec}).  
For the north component we have fitted a 3-T plasma, with temperatures in the range between 6 and 40 MK. 
The emission is dominated by the softer plasma with temperatures in the range between 6 and 13 MK.
The average temperature was defined as $\log T_{\rm av}= (\Sigma_i \log T_i \times \rm{EM_i})/\rm{EM_{total}}$, leading to $T_{\rm av}=11.3$~MK. 
The upper limit for the component at 40 MK was not well constrained.

For the southern star we could not fit the soft component at 6 MK that was obtained for northern star.
We have found a 2-T plasma with temperatures 8 and 34 MK, the emission is dominated by the plasma at 8 MK. 
Again in this case, the upper limit for the highest temperature (34 MK) was not well constrained.
The average temperature was 12.2 MK, slightly higher than the average of the northern star.

We remark that the sum of the emission measures, as well as the luminosity, obtained for both components of the binary does not match the value of the total emission measure found in our fits to the combined spectrum of the binary (see \ref{binaryspec}) due to the different extraction radii used for the binary and their components.

\subsubsection{Flare}\label{flare}

To analyze the spectrum from the southern star during the flare we have extracted only the zeroth order spectrum, since we do not have enough signal in the gratings.
We have fitted a multi-$T$ model leaving the photoelectric absorption component, abundances and temperatures fixed to the values from the best-fit to the quiescent state of the southern star. 
We have then added an extra temperature component for which we have left only the emission measure and temperature free to vary. 
We could not add more than one temperature component to our model, probably due to the low signal obtained.
 
Our fit shows that the flare emission is dominated by a plasma at $\sim 30$~MK (see Table~\ref{table:flare}).
The luminosity during the flare increases to $7.6\times 10^{30}$~erg s$^{-1}$, which is a factor of five higher than the luminosity of the quiescent state. 
The high temperature found during the flare is also present in the quiescent state, indicating that the variable component during the flare originates mainly in the hotter part of the emission measure distribution.

\subsubsection{HDE~245059 Binary}\label{binaryspec}

The combined zeroth order CCD spectrum of HDE~245059 was extracted after merging the event files from the three observation epochs following standard CIAO 
procedures. 
This was possible thanks to the similar pointing and setup between the observations. 
We have used a circular region with 3\farcs9 radius. 
The background was extracted from an annular region with radii of 4\farcs6 and 21\farcs \,
We considered events at energies within the range $0.2 - 7$~keV.

\begin{figure*}
\begin{center}
\begin{tabular}{cc}
\includegraphics[angle=0,scale=0.22]{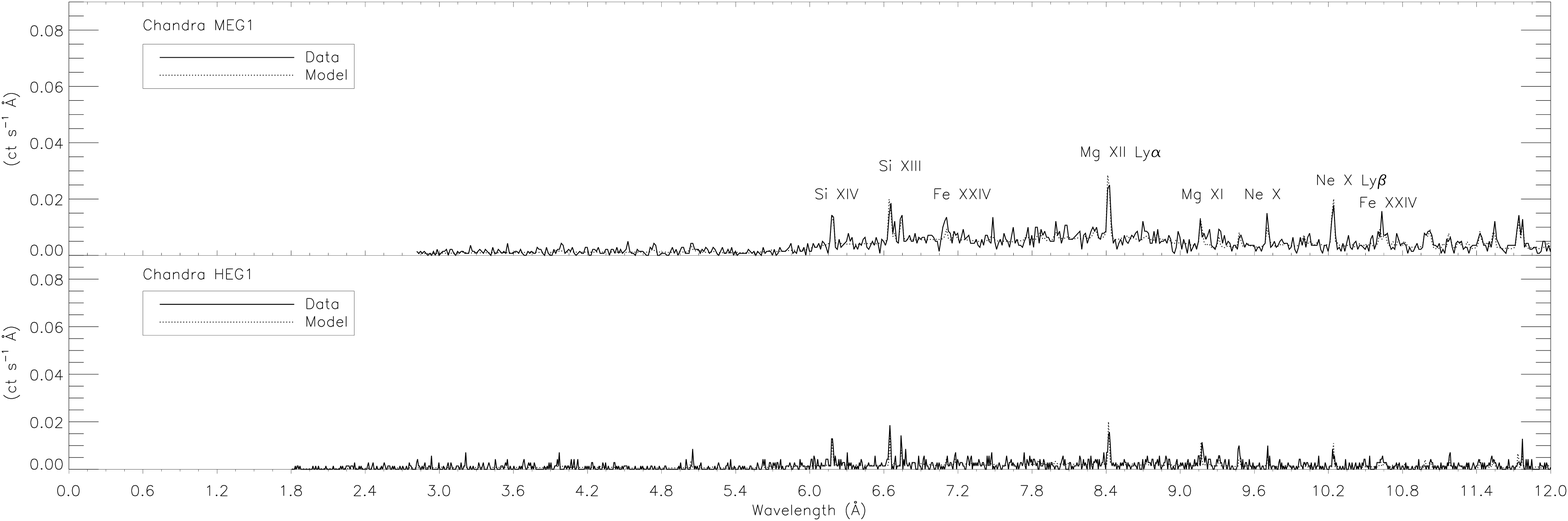} \\
\includegraphics[angle=0,scale=0.22]{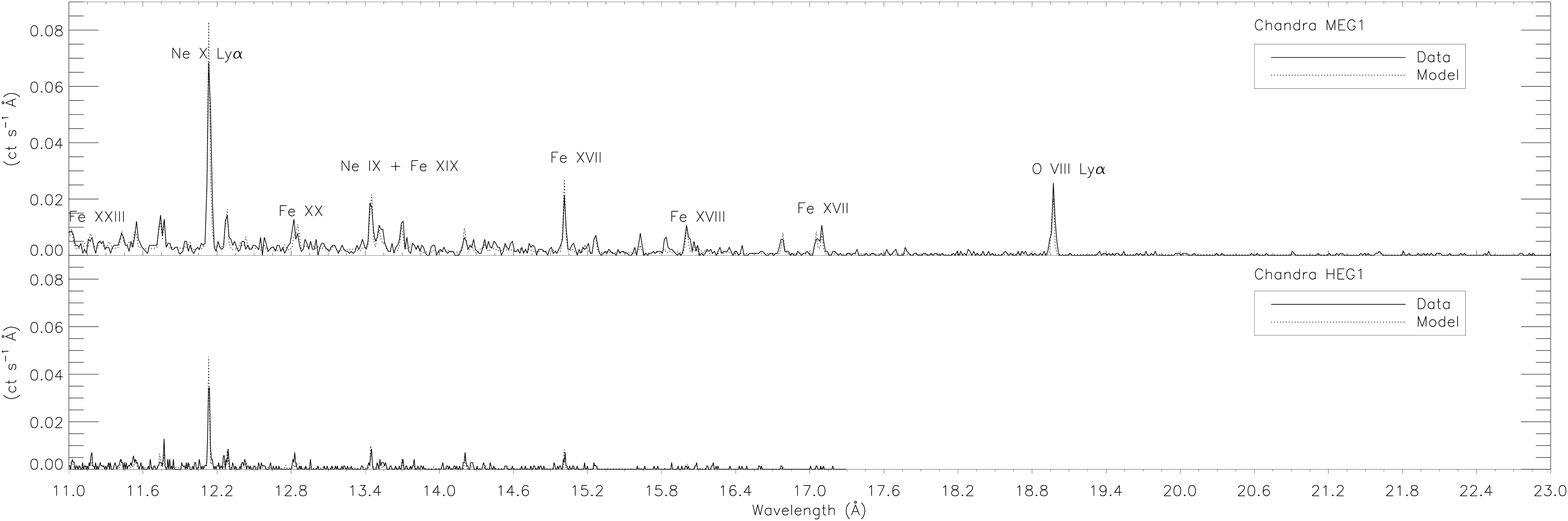}
\end{tabular}
\end{center}
\caption{HEG $\pm 1$ and MEG $\pm 1$ average spectra and best-fit model overlaid for the binary. Bright lines are labeled. 
The spectra were binned by a factor of 3.}
\label{fig:spect+model}
\end{figure*}

The grating spectra of the HDE~245059 binary were extracted for both the HEG (High Energy Grating, 0.8 - 10 keV) and MEG (Medium Energy Grating, 0.4 - 5 keV) orders $\pm1$.
We have selected the wavelength range to 2.8 $-$ 25 and 1.8 $-$ 17.3 \AA~ for MEG and HEG, respectively.
The spectra were binned by a factor 3 in wavelength, in particular to sum the contribution of both stars in the MEG spectra (since both stars are aligned with the dispersion direction). 
The respective spectral responses were generated using standard CIAO tools.
The spectra of all the observation epochs were then merged using the CIAO procedure {\it merge\_all}.
In Figure \ref{fig:spect+model} we show the binary grating spectrum for MEG and HEG first orders. We have labeled some important emission lines.

We have analyzed the spectrum of the binary HDE~245059 during the quiescent state using the combination of the zeroth order spectrum plus the grating spectrum from HEG order 1, and MEG order 1.
For the fits we have used three methods: 
a multi-$T$ component model as a discretization of the EMD, a continuous EMD form Chebychev polynomials, and a continuous EMD approximated by two power laws.
For the three methods we have obtained abundances, emission measures, and an estimate of the photoelectric absorption component parameterized by the hydrogen column density.
In all cases we have used the combination of the summed zeroth order, HEG$\pm 1$ and MEG$\pm 1$ spectra.

\subsubsection{Method 1: Discrete Emission Measure Distribution with a Multi-temperature Model}
\label{discrete}

This method uses a discretization of the EMD by a multi-temperature VAPEC model, which calculates both line and continuum emissivities for a hot, optically thin plasma. 
We have combined this model with a photoelectric absorption component (WABS), parametrized by the neutral hydrogen column density $N_{\rm H}$. 
Reference abundances were set to solar photospheric values \citep{grev98}. 
We have used several isothermal components for the plasma emission. 
A 4-temperature plasma model gave the best fit compared to models with additional or fewer plasma components.
Our results are summarized in Table~\ref{table:fits-total}. 
All the values are displayed with their respective errors calculated for a confidence range of 68\%. 
We have found a plasma between 4 and 50~MK. 
The average temperature was defined as $\log T_{\rm av}= (\Sigma_i \log T_i \times \rm{EM_i})/\rm{EM_{total}}$, leading to $T_{\rm av}=10.7$~MK.
The value of the hydrogen column is very low, $N_{\rm H} = 8\times 10^{19}$~cm$^{-2}$, corresponding to $A_V = 0.04$~mag, which is consistent
with the 2MASS colors of the binary ($J-K=0.6$) and with photospheric colors of early K type stars. 
We remark that the results obtained from the fit to the quiescent state are not different from the results obtained from fitting the total average spectrum (considering also the flare).

The best-fit model overlaid to the HEG$\pm 1$ and MEG$\pm 1$ data for the quiescent state are shown in Figure~\ref{fig:spect+model}. 
The brightest line detected is Ne~X~Ly$\alpha$ at 12.13~\AA. 
Significant emission comes also from O~VIII~Ly$\alpha$ at 18.97~\AA, Ne~X~Ly$\beta$ at 10.23~\AA. 
Interestingly there is no detection of O~VII or N~VII despite the low $N_{\rm H}$.
Lines from highly ionized states of Fe~XVII to Fe~XXIV confirm the presence of a wide range of plasma temperatures revealed by our fits. 

\subsubsection{Method 2: Continuous EMD from Chebyshev Polynomials}\label{cheby} 

The differential EMD $\varphi (T)$ is given by 

\begin{equation}
\varphi (T) = n_{\rm H}n_{\rm e} \frac{dV}{dT}   \,\,\,\,\,\,\,\,\,\,\,\,     \rm{(cm^{-3}K^{-1})}
\end{equation}

where the total EM is given by EM$_{\rm tot}$ = $\int \varphi (T) T \Delta {\rm log} T$ = $\int \varphi (T)T d({\rm ln} T)$.
A graphical representation of the EMD independent of the grid bin size ($\Delta {\rm log} T$) is given by EMD($T$)/$\Delta {\rm log} T$.
This method assumes that the shape of $\varphi (T)$ can be approximated by the exponential of a polynomial given by $\varphi (T)=\alpha e^{\omega(T)}$, where $\alpha$ is a normalization constant and $\omega(T)$ is a polynomial function of the temperature, which we have chosen to be a Chebyshev polynomial (see Lemen et al. 1989, Audard et al. 2004).

Our model uses a grid of temperatures in the range log$T ({\rm K})=8 - 10$  with $\Delta {\rm log} T=0.2$ dex for a polynomial degree of $n=8$ which gives the optimal fit.
Coronal abundances and the photoelectric absorption were left free to vary. 
The results obtained with this method are displayed in Table \ref{table:chebychev}.

\subsubsection{Method 3: Continuous EMD approximated by two power laws} \label{powerlaw}

This method is based on a continuous EMD described by two power laws; one at low temperatures with slope $\alpha$, and one at high temperatures with slope 
$\beta$ (see Telleschi et al. 2007b). 
The EMD peaks at the temperature $T_0$ which also represents the limit between the two regimes.

\[Q(T) = \left\{ 
\begin{array}{l l}
  EM_0(T/T_0)^{\alpha} & \quad \mbox{ for $T \le T_0$}\\
  EM_0(T/T_0)^{\beta} & \quad \mbox{for $T > T_0$}\\
\end{array} \right .\]

The normalization parameter is defined as the EM in the temperature bin at $T_0$.
The free parameters are: $T_0$, EM$_0$, $\alpha$, $\beta$, $N_{\rm H}$, and the elemental abundances. 
We have used 2 approximations; first leaving $\alpha$ and $\beta$ free to vary, and after fixing the value of $\alpha=2$ leaving $\beta$ free to vary. 
The value $\alpha=2$ is usually found in magnetically active main-sequence and pre-main sequence stars. 
$\Delta$log$T$=0.1 dex was set in both cases. 
The results are presented in Table \ref{table:PL}.

\begin{figure}[hbp]
\begin{center}
\includegraphics[angle=0,scale=0.4]{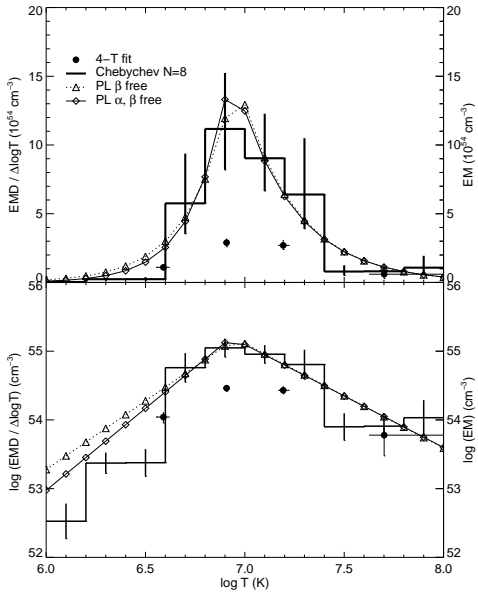}
\end{center}
\caption{Emission measure distribution of the binary HDE~245059 spectra obtained from the different methods described in the text. 
We have plotted the emission measure per bin; EMD($T$)/$\Delta$ log $T$, which is independent of the bin size allowing comparison with the other methods. 
In the case of the multi-$T$ approach we have plotted the EM for each component(filled circles).
For the upper panel we have used a linear scale in order to emphasize the dominant plasma at ~10 MK.
In the lower panel we have used logarithmic scale to show the presence of weaker components of plasma at very high temperatures. }
\label{fig:EMD}
\end{figure}

In Figure \ref{fig:EMD} we have plotted the EMD of the binary obtained by the different methods applied.
In order to compare them we have used the quantity EMD($T$)/$\Delta$log$T$ which is independent of the bin size. 
The total volume emission measure obtained from the different methods are similar: 7.3 for method 1, 7.1 for method 2, 7.5 for method 3 $\alpha$ fixed and $\beta$ free, and 7.3 for method 3 with $\alpha$ and $\beta$ free (all values are in units of 10$^{54}$ cm$^{-3}$). 

\subsection{Ftting}\label{singlefits}
\subsubsection{Electron Densities}\label{density}

\begin{figure}[hbp]
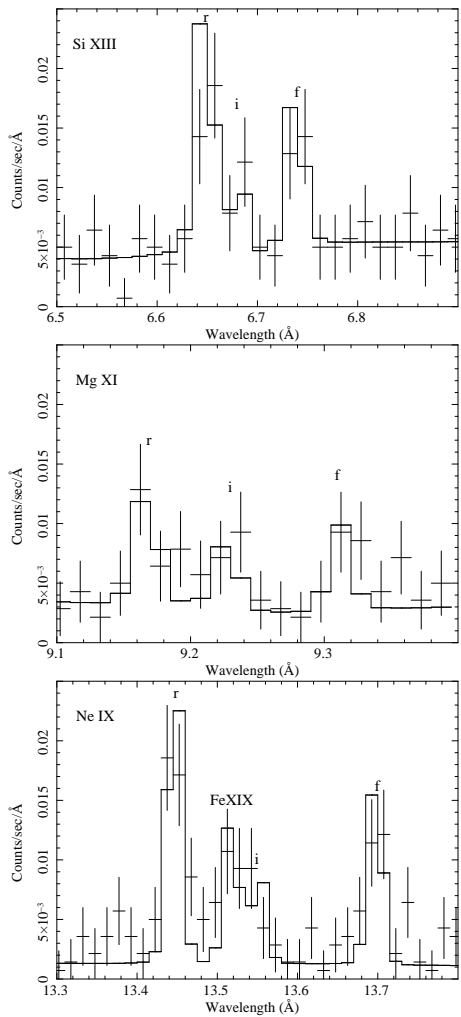

\begin{center}
\includegraphics[angle=270,scale=0.25]{SiXIII.ps}
\includegraphics[angle=270,scale=0.25]{MgXI.ps} 
\includegraphics[angle=270,scale=0.25]{NeIX.ps}
\end{center}
\caption{He-like triplets in the grating spectra. We have plotted the MEG spectrum of the binary and in histogram the delta profiles used to fit the resonance
($r$), intercombination ($i$), and forbidden ($f$) lines. For Ne~IX we have also added the blend with Fe~XIX.}
\label{fig:he-like}
\end{figure}

We have obtained the fluxes of the density sensitive He-like triplets Si~XIII, Mg~XI, and Ne~IX from the binary grating spectrum (HEG and MEG first order).
The first step was to fit the continuum to get an estimate of the plasma temperature. 
We used a bremsstrahlung model, taking the spectrum without the contribution of the bright lines and considering the wavelength range 
between $1.2$ and $40$~\AA. 
We have obtained a plasma temperature of $12.4 \pm 0.3$, MK close to the average temperature found in our previous best-fit of the grating plus zeroth order
spectrum. 
To fit the triplet lines (resonance ($r$), intercombination ($i$), and forbidden ($f$)) we used the combination of delta functions for the line profiles and the above mentioned 
bremsstrahlung model for the continuum. 
We have fixed the continuum parameters and fitted the triplets only in the wavelength range of interest.
The only free parameters in our fits were the line fluxes, since the wavelength of each line was fixed.
Since the Ne~IX triplet is blended with Fe~XIX, we added a delta profile to consider this blend.
The results are summarized in Table~\ref{table:helikebinary}.
In Figure \ref{fig:he-like} we have plotted the three triplets for the MEG first order spectrum including the delta profiles used to fit the lines.
Based on theoretical models \citep{porquet01}, using the $R=f/i$ ratio and the temperature of the emitting plasma we have obtained an estimate of the plasma electron density. 
We have calculated the densities using confidence levels of 68 and 90\% (see Table \ref{table:densities}). 
For the 68 \% confidence level, we have obtained $n_{\rm e} = 5 \pm 5 \times 10^{11}$~cm$^{-3}$ from the Ne~IX triplet, 
 $n_{\rm e} = 1 ^{+3} _{-0.5} \times 10^{13}$~cm$^{-3}$ from Mg~XI, 
and $n_{\rm e} < 5 \times 10^{13} $~cm$^{-3}$ from Si~XIII, which is consistent with the low density plasma.
For the 90 \% confidence level, we have obtained $n_{\rm e} = < 2.0 \times 10^{12}$~cm$^{-3}$ from the Ne~IX triplet, 
 $n_{\rm e} = 1 ^{+6} _{-0.8} \times 10^{13}$~cm$^{-3}$ from Mg~XI, 
and $n_{\rm e} < 3 \times 10^{14} $~cm$^{-3}$ from Si~XIII.
Therefore, we prefer to err on the safe side and conclude that there is no evidence of high densities in HDE~245059.

\subsubsection{Lines fluxes from each binary component}\label{singlelines}

\begin{figure*}
\begin{center}
\begin{tabular}{cc}
\includegraphics[angle=0,scale=0.2]{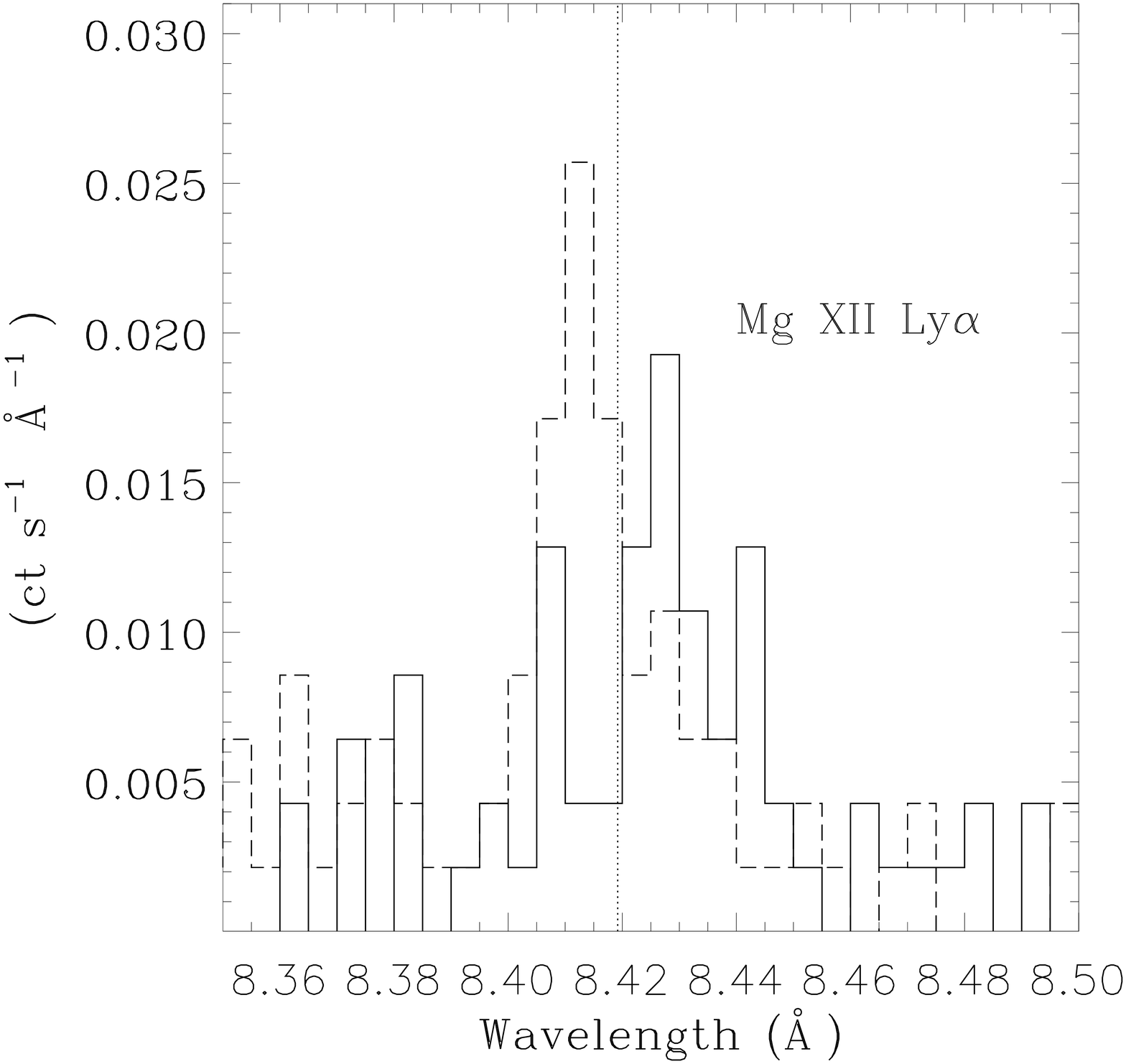}
\includegraphics[angle=0,scale=0.2]{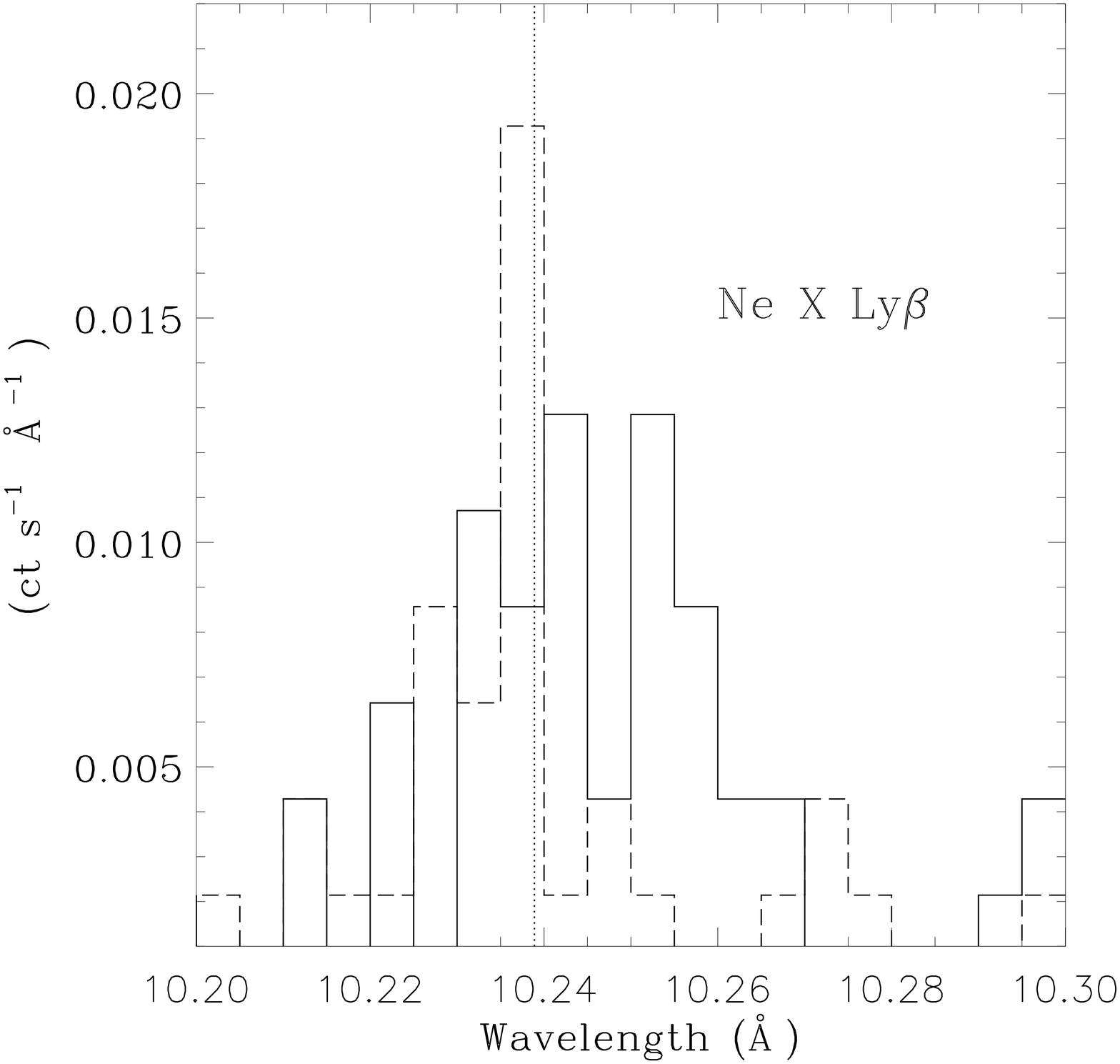} \\
\includegraphics[angle=0,scale=0.2]{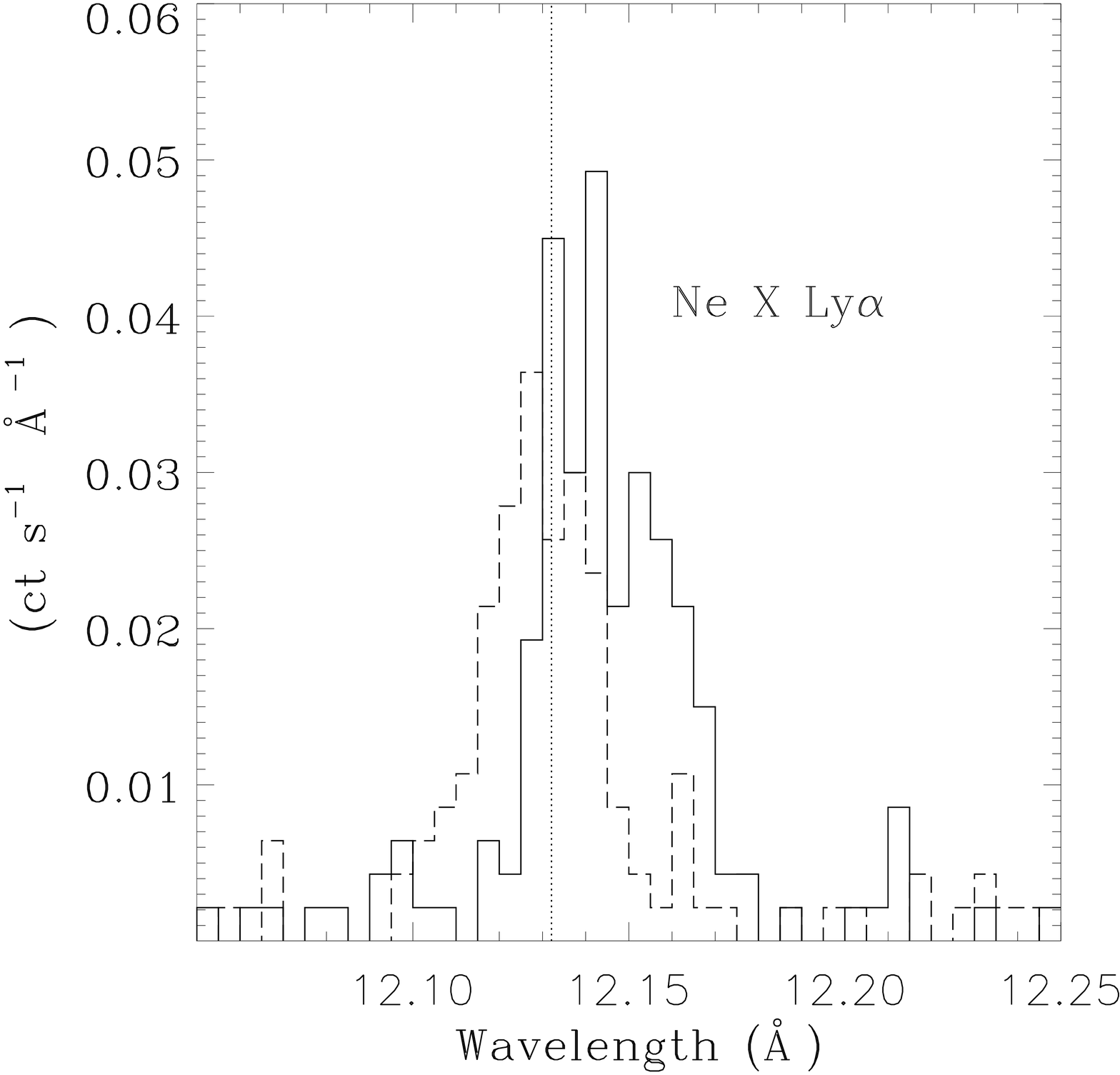} 
\includegraphics[angle=0,scale=0.2]{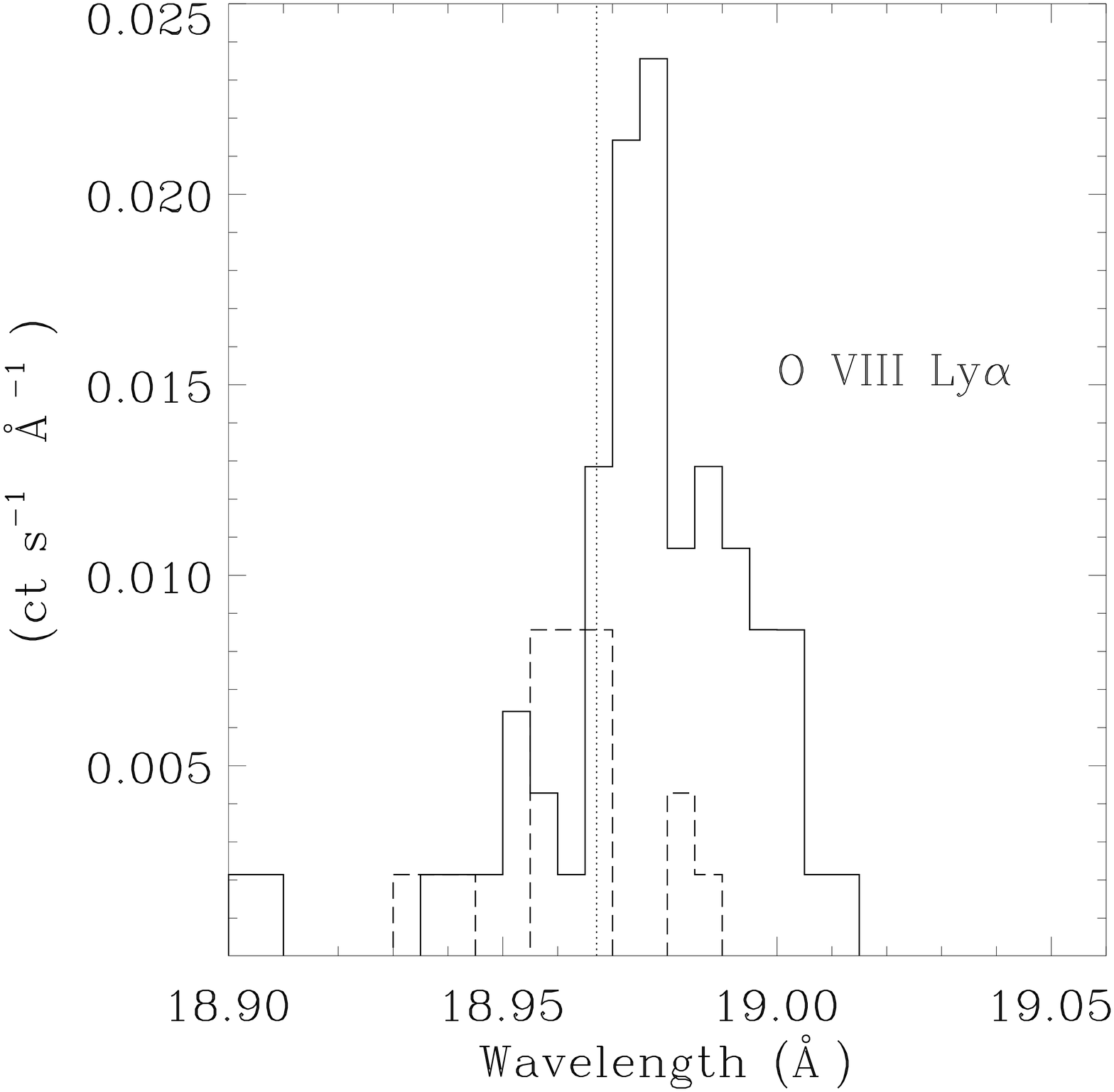}
\end{tabular}
\end{center}
\caption{The plot shows the Mg~XII~Ly$\alpha$, Ne~X~Ly$\alpha$, Ne~X~Ly$\beta$, and O~VIII~Ly$\alpha$ lines for the grating data. 
In solid line the MEG-1 and in dashed line the MEG+1.
The line were modeled with a bremsstrahlung plus a gaussian model taking into account the contribution of both stars.
We considered the profiles to be purely instrumental. 
}
\label{fig:OVIII}
\end{figure*}

We have also attempted to obtain individual fluxes from several lines for each component of the binary. 
We have used only the first order spectra of MEG because of the higher signal to noise, and because the binary was aligned along the MEG dispertion direction, 
allowing us to disentangle the two components in wavelength space.
The spectrum was re-extracted using a bin size of 0.015 \AA. 
To fit the single lines we used the method described in section \ref{density}; a combination of a delta function for the line profile and the bremsstrahlung model for the continuum.
The fits to the MEG$+1$ and MEG$-1$ data were made simultaneously because the shift of the lines will occur in opposite directions from 
the origin of the reference wavelength. 
For each line we considered only the instrumental profile, therefore we had 4 delta profiles: 
2 for the MEG$+1$, northern and southern star and the same for the MEG$-1$.
From the grating equation, we calculated the shift of the lines.
The line shifts were smaller than the line spread function full width at half maximum (FWHM), with a typical separation of $~ 17$ m\AA.  
To fit the lines, we fixed the energy at which they were expected to be found, leaving only the fluxes free to vary.
Besides, the fluxes from each star were linked for MEG$+1$ and MEG$-1$.
The fits thus considered 2 free parameters: the flux of the line from each star.
The calculations were made whenever the signal for a given element was high enough; we have then obtained fluxes for O~VIII Ly$\alpha$, Ne~X~Ly$\alpha$, Ne~X~Ly$\beta$, and Mg~XII~Ly$\alpha$. 
For the He-like triplets the signal was too low, thus we did not obtain fluxes for the single binary components.
The results are displayed in Table \ref{table:NS_flux},
in Figure \ref{fig:OVIII} we have plotted the line profiles for the full resolution grating MEG$\pm$1 data.

\section{Discusion}%
\subsection{Environment and Binarity}\label{env-bin}

The absence of a disk in the WTTS HDE~245059 members despite their young age might be closely related to the environment where this young binary has been formed.
The evolution of the region around $\lambda$ Ori has been discussed in detail \citep{dolan99,dolan01,dolan02,barrado07}. 
Dolan~\&~Mathieu (1999, 2001, 2002) presented the hypothesis of a supernova explosion that cleared out the region about 1~Myr ago leaving a molecular ring.
During the phase prior to the supernova explosion the stars in the cluster would have been confined in a small 
region and the circumstellar disks cleared away by photo-evaporation.
This hypothesis is supported by the small fraction of CTTS in the region, $\sim 7~\%$ \citep{dolan99}, 
which is low when compared with clusters with similar properties. 
\citet{barrado07} have studied the members of the $\lambda$ Ori with the \emph{Spitzer} satellite, finding a $31 \%$ of members with disks in $\lambda$ Ori, but only $14 \%$ with thick disks. According to this study, the presence of CTTS near the center of the cluster would suggest that massive stars and the supernova explosion had no major effect on the disks, or that they have been formed after the explosion.
Recent high resolution optical spectroscopy of the $\lambda$~Ori cluster \citep{sacco08} has given a fraction of stars with disks of $ 28 \%$, higher than the previous studies.
Our new estimates of the properties of the HDE~245059 binary (see section \ref{stellar properties}) give an age of $\approx 2-3$~Myr: at this age the binary would have experienced the supernova explosion, according to Dolan~\&~Mathieu (1999, 2001, 2002) hypothesis. 
Another aspect to be considered is the influence of binarity in the disk truncation.
This hypothesis has been discussed by \citet{kraus08}, who presented recent results
showing that several of the young stars without disks in a survey of nearby star-forming regions are close binaries. 


\subsection{Stellar properties}
\label{stellar properties}

The discovery of the binarity of HD 245059 prompts us to re-evaluate
the stellar properties of the system. 
The few physical properties; mass, age, spectral type, and effective temperature were previously obtained under the assumption of a single star
\citep{padgett96,stone85}. 
Our near-infrared and X-ray data have been the first to separate it into a binary. 
The question that arises immediately is how different the properties of the HDE~245059 members are. 
The magnitude differences from the near-infrared images in the $H$, $K$, and Br$\gamma$ bands 
suggest that both components have similar colors, 
but the northern star is brighter in both near-infrared and X-ray images, which might indicate that is the more massive of the system. 
Throughout this analysis, we assume a distance of 400\,pc to the system. 
We further assume that the flux we receive from
each component can be entirely attributed to the stellar photosphere,
without contribution from accretion for instance, in line with the WTTS
status of the system.

In a first approach, the absolute near-infrared magnitudes of each
component can be used to estimate its main properties (mass and
age). Combining the unresolved 2MASS photometry of the system with our
new $H$ and $K$ band flux ratios, we determine absolute magnitudes of
$M_H=-0.07$ and $M_K=-0.20$ for the primary and $M_H=+0.91$ and $M_K=+0.68$ for
the secondary\footnote{Note that extinction, if present in front of the system, would yield even higher absolute magnitudes.}. 
Such absolute magnitudes are too bright for solar-like stars even at ages
as young as 1\,Myr, based on the stellar evolutionary models of
\citet{baraffe98} and \citet{siess00}. To explore higher-mass
regimes, we adopt the models of \citet{siess00} which extend up to
7\,$M_\odot$. Based on this model, for ages of 4\,Myr or more, only
B-type stars reach the observed brightness of HD 245059 north and south. 
At 3\,Myr, at least the primary would have to be a B star. Since this can
be confidently excluded from the spectral analysis of \citet{padgett96},
we conclude that the system is no older than 2\,Myr. Assuming stellar
ages in the 1--2\,Myr range (since star formation appears to have
ceased about 1\,Myr ago, see e.g. \citet{dolan01}), a primary of
2.5--3.5$\,M_\odot$ and a secondary of 2--3$\,M_\odot$ would match the
observed near-infrared magnitudes of the two components.

To improve on this estimate and take advantage of a broader dataset,
we perform a fit to the optical and infrared SED of the system. For
this purpose, we use the unresolved $UBV$ photometry from \citet{stone85}, 
the unresolved $J$ magnitude from 2MASS, the spatially
resolved photometry derived above and the 3.6, 4.8, 5.6 and 8\,$\mu$m
unresolved IRAC photometry determined from archival images and using
default recipes for aperture correction around point sources. This
represent a total of 12 independent measurements. We used a grid of
{\it NextGen} stellar spectra from \citet{baraffe98} with $\log g
=4.0$ (our results are largely insensitive to the assumed surface
gravity). We used 5 free parameters in our model: the stellar
effective temperatures $T_{\rm eff}(N)$ and $T_{\rm eff}(S)$, the stellar
radii $R_N$ and $R_S$, and the extinction $A_V$. To ensure that we did
not miss the best possible solution, we conservatively allowed for
wide ranges of initial guesses (4600--7600\,K for $T_{\rm eff}(N)$,
2400--7000\,K for $T_{\rm eff}(S)$, 3.5--6\,$R_\odot$ for $R_N$,
2--9\,$R_\odot$ for $R_S$ and 0--1.5\,mag for $A_V$). In this
procedure, we do not use information from evolutionary models as prior
and only require that $T_{\rm eff}(S) \leq T_{\rm eff}(N)$. Overall, we tested
about 5 million independent combinations of the 5 free parameters,
estimating a reduced $\chi^2$ value for each model. We then used a
Bayesian inference method to explore the parameter space: each model
in the grid is assigned a probability $p=e^{-\chi^2/2}$, and 1- and
2-dimensional probability distributions for individual and pairs of
free parameters are then produced by marginalizing the hypercube
against the other dimension.

\begin{figure}[hbp]
\begin{center}
\includegraphics[angle=0,scale=0.5]{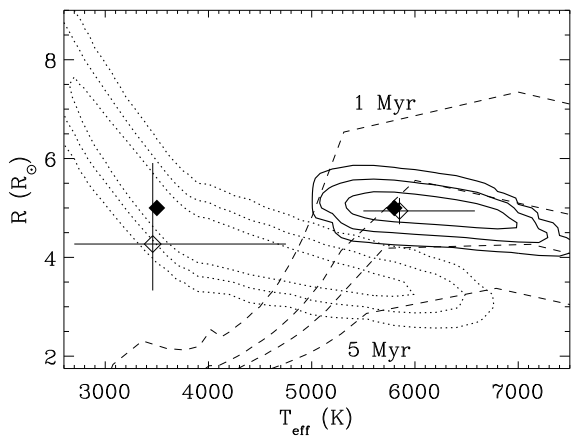}
\end{center}
\caption{Estimated stellar radii and effective temperatures
based on our fit to the SED. Solid and dotted contours represent the
68.3, 95.5 and 99.7\% confidence levels for the primary and secondary,
respectively. Filled diamonds indicate the best fitting stellar models
in our grid (corresponding to $A_V=0.7$\,mag). Open diamonds represent
the most probable models from the mode of the 1-dimensional
probability distributions in our Bayesian approach; associated
uncertainties represent the 68.3\% confidence level intervals around
these models. From top to bottom, the dashed lines represent the 1, 2,
3 and 5\,Myr isochrones from the evolutionary model of \citep{siess00}. 
The available dataset is consistent with a 2--3\,Myr age
for the system (see text for more detail). 
}
\label{fig:hde245059_SED}
\end{figure}

Using the mode of each 1-dimensional probability distribution and
defining a 68\% confidence level interval around it, we infer
$T_{\rm eff}(N) = 5850^{+730}_{-350}$\,K, $T_{\rm eff}(S) =
3460^{+1290}_{-760}$\,K, $R_N = 4.94^{+0.26}_{-0.27}\,R_\odot$ and
$R_S = 4.27^{+1.64}_{-0.94}\,R_\odot$. However, as illustrated in
Figure~\ref{fig:hde245059_SED}, there is a substantial ambiguity between $T_{\rm eff} (S)$
and $R_S$ due to the fact that we only have 2 measurements that
directly constrain the secondary. A cool but large secondary fits
equally well the data than a warmer and more compact star. This
correlation between the stellar parameters also explains the
difference between the absolute best model (i.e., lowest $\chi^2$) and
the most probable stellar parameters based on the 1-dimensional
probability distributions. Both estimates nonetheless agree within the
uncertainty, although we prefer to use the 2-dimensional probability
distributions to estimate the stellar properties.

The final step in this analysis consists in overplotting the
predictions of the \citet{siess00} evolutionary models. We note
that both the stellar luminosity and radius are direct output of these
models so this is a natural set of parameters to compare model and
data. We overplot in Figure~\ref{fig:hde245059_SED} the 1, 2, 3 and 5\,Myr isochrones and
readily conclude that the primary star is most likely
$\approx2$\,Myr-old. Similar to the conclusion based solely on the
near-infrared magnitudes, stellar ages beyond 3\,Myr can be
confidently excluded. While the best fit to the secondary suggests an
age even younger than 1\,Myr, the 2\,Myr isochrone does intercept the
68\,\% confidence level contour. Using 6000\,K and 5$\,R_\odot$
(equivalently, 29\,$L_\odot$) for the primary, interpolation in the
\citet{siess00} model yields a stellar mass of 2.7\,$M_\odot$ and
an age of 2.7\,Myr. From the extent of the 68\% confidence level
contour, we estimate uncertainties of 0.5\,$M_\odot$ and 1\,Myr. 
Based on the 2.7\,Myr best-fitting age, we further infer $M_S = 2.3$--2.4\,$M_\odot$.

While relatively large uncertainties remain due to the limited number
of resolved photometric measurements of the system, we conclude that
the stellar masses are $\approx3\,M_\odot$ and $\approx2.5\,M_\odot$
and the age of the system is $\approx3$\,Myr. High spatial resolution
optical data, which are currently unavailable for the system, would
greatly improve the accuracy on these parameters. Nonetheless, we
consider that this is a significant improvement over the previous
estimates that did not take into account the fact that the system is
indeed a binary. As a final note regarding the stellar properties, we
note that the age of HD\,245059 is younger than the average age of
other members of the $\lambda$\,Ori association but also consistent
with that of the youngest systems in the associations.

\subsection{Spectral X-ray Properties}\label{spect-prop}

The fits to the high-resolution grating spectroscopy data have revealed that both stars have similar spectral properties and emission measure distributions.
Indeed, the spectra of the single components are consistent with the average spectrum of the binary.
The similar temperatures of both stars allowed us to fit easily the grating spectra together.

We do not detect the oxygen triplet in HDE~245059, either in the average or in the single spectra, despite the low $N\rm{_H}$ value.
This triplet is used as an electron density indicator in the cool plasma component, expected to be present in the spectra of accreting stars.
On the other hand we do find a soft plasma component at 3.8~MK in the average spectrum:  
the He-like triplet of Ne~IX was detected with low signal to noise and it was found to be blended with the Fe~XIX line. 
We also detect emission from Fe~XVII blended with other lines. 

We have estimated an upper limit flux for the O~VII triplet (see Table~\ref{table:helikebinary}) at 2 confidence levels: 68~\% and 90~\% for the average binary spectrum.

In order to determine the nature of our non-detection of the O~VII triplet,
we have fitted some single lines for the combined spectrum of the HDE~245059 binary: O~VIII~Ly$\alpha$, Ne~IX, Ne~X~Ly$\alpha$, and Fe~XVII at 15~\AA\ (Table \ref{table:helikebinary}).
We have compared the results with the observed line fluxes of 3 young active stars: 47 Cas B, EK Dra \citep{telleschi05} and AB Dor \citep{garcia08}.
We have found the line fluxes of the HDE~245059 binary to be about a factor of 100 larger than the fluxes of the comparison stars.
Using this ratio for the O~VII resonance line at 90\% confidence level, our upper limit for HDE 245059, $ < 2 \times 10^{-14}$ (ergs cm$^{-2}$ s$^{-1}$), remains close to what could have been expected
based on the comparison stars. We conclude that the non-detection of the O~VII triplet is probably due to a lack of sensitivity around 22~\AA.

\begin{figure}[hbp]
\begin{center}
\includegraphics[angle=90,scale=0.3]{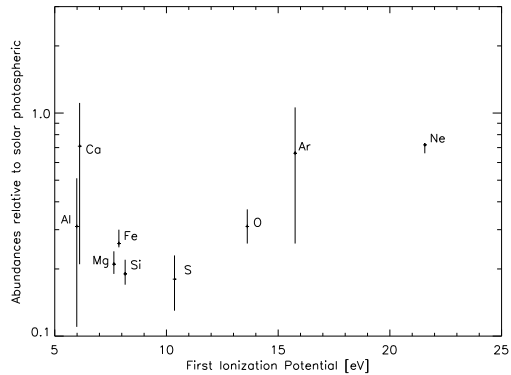}
\end{center}
\caption{Coronal abundances with respect to the solar photospheric values, obtained by fitting the combined zeroth order, HEG, and MEG data of the binary.
The plot shows the best-fit values using 4 isothermal plasma components plus interstellar absorption. We see evidence of an inverse FIP effect}
\label{fig:fip}
\end{figure}

Observations of the Sun show that abundances are related to the first ionization potential (FIP) of the elements in such a way that elements with low FIP ($\leq 10$ eV) 
are overabundant in the solar corona when compared to the photosphere and high FIP elements ($\geq 10$ eV) have similar abundances in the corona and the 
photosphere \citep{feldman92}.
Observations of young, magnetically active stars show that the FIP effect is inversed with respect to the Sun, i.e. low FIP elements are underabundant 
in the stellar coronae relative to elements with high FIP \citep{brinkman01,audard03}. 
Figure \ref{fig:fip} shows the coronal abundances with respect to the solar photospheric values for our binary plotted against the FIP. 
The abundances follow the trend of an inverse FIP effect, except for the Ca abundance which has a higher value, 
but we recall that the uncertainty in Ca abundance from our results is also high.

A common problem found in the studies of the coronal element abundance is the lack of measurements of the stellar photospheric abundances, 
as they are difficult to obtain due to the large rotational velocity of magnetically active stars. 
Furthermore, a good knowledge of the stellar parameters and atmospheric models are also needed.
To sort out this problem,  the stellar photospheric abundances are used as a reference set.
In the case of HDE~245059 \citet{padgett96} obtained the photospheric iron abundance $\rm{[Fe/H]}=-0.07 \pm 0.13$ using as reference set the solar photospheric abundances from \citet{grev84}.
Using the revised reference set of \citet{grev98}, the stellar photospheric abundance is $[\rm{Fe/H}]=+0.10$.
Our coronal abundances have been obtained from the best fit to the zeroth order, MEG$\pm1$, and HEG$\pm1$ using as reference set the solar photospheric values from \citet{grev98}. We have obtained an iron abundance of $\rm{[Fe/H]}=-0.64$.
Thus, the difference between the photospheric and coronal values is $[\rm{Fe/H}]_{\rm photospheric}-\rm{[Fe/H]}_{\rm coronal}=+0.74$, i.e., the coronal iron abundance is 5.5 lower than the photospheric one.
This result is consistent with the inverse FIP effect expected in young magnetically active stars.

\section{Summary \& Conclusions}

We have obtained the \emph{Chandra} high resolution spectrum for HDE~245059.
Thanks to our X-ray and near-infrared data we have resolved this X-ray luminous WTTS to be a binary.
We have attempted to get an estimate of the properties of the single components of the HDE~245059 binary based in the combination of the infrared magnitude differences,
evolutionary models and SED determination.
Our analysis gave a system of $\approx2-3$\,Myr with masses $M_{\rm N} \approx 3 M_{\odot}$ and $M_{\rm S} \approx 2.5 M_{\odot}$, rather high values for low-mass 
pre-main sequence stars. 
This is the first attempt to constrain the binary component masses, further studies are needed to obtain more accurate estimates.

In the X-rays we were able to resolve both binary components in the zeroth order image and in the grating spectra.
We have analyzed the zeroth order spectrum for each component of the binary using a multi-$T$ plasma model.
For the northern star we have found a temperature between $\sim6$ to $\sim 40$~MK, the dominating component being a plasma between $\sim6$ and $\sim13$~MK. For the southern star, we have found temperatures between $\sim 8$ and $\sim34$~MK.

During our observations, split in three runs, we have detected a flare from the southern star, the fainter of the system in average.
An analysis of the zeroth order spectrum of this star during the flare has gave a plasma with a temperature of $\sim30$~MK, consistent with the high temperature component obtained for the quiescent state, and a luminosity higher by a factor of $\sim 5$ when compared with the quiescent state.

We have derived the properties of the plasma from the combined zeroth order, MEG$\pm1$, and HEG$\pm1$ data for the binary during the quiescent state.
We have analyzed the spectrum using 3 methods: a continuous emission measure distribution from Chebyshev polynomials,
a continuous emission measure distribution approximated by two power laws, and the classical discretization of the emission measure distribution by a multi-$T$ optically thin plasma model.
The three methods are consistent and we have found the emission to be dominated by a plasma in the temperature range 8~-~16 MK, with a low value for the column density, N$_{\rm H} = 8 \times 10^{19}$~cm$^{-2}$. 
The low N$_{\rm H}$ might be due to the clearing of the inner region of $\lambda$ Ori due to a supernova event.
The X-ray luminosity of the binary is high, L$_{\rm X} \sim 10^{31}$ $\rm{ergs~s^{-1}}$. 

The abundance pattern shows an inverse FIP effect, except for Ca which could not be well constrained by our fits.
This result was confirmed for Fe by comparison of the coronal abundance from this work with the photospheric value from \citet{padgett96}.
Despite the low N$_{\rm H}$, we did not detect the density sensitive O~VII triplet,
but we did obtain an estimate of the flux upper limit for this triplet from the binary grating spectrum. 
We have compared the fluxes obtained from our fits to O~VIII, Ne~IX, Ne~X,
and Fe~XVII with observed fluxes of three active stars. 
Our upper limit flux for the O~VII triplet is consistent with the flux differences when compared with the active stars.
Based on the 90\% confidence range, we have obtained upper limits to the plasma electron densities from He-like triplets; $n_{\rm e } < 2 \times 10 ^{12}$~cm$^{-3}$ for Ne~IX, $< 6 \times 10^{13}$~cm$^{-3}$ for Mg XI, and $< 3 \times 10^{14}$~cm$^{-3}$ for Si~XIII. 

According to our analysis, the properties of the HDE~245059 binary in the X-rays are similar to what is observed in other WTTS. 
Its accretion history might have been affected by the clearing of the region surrounding $\lambda$ Ori, probably by a supernova explosion, but its coronal properties have not been strongly modified.

\acknowledgments

M. Audard and C. Baldovin Saavedra acknowledge the support of the Swiss National Science Fundation grant PP002-110504. 
Support from the \emph{Chandra} award SAO GO5-6012X is also acknowledged.
We would like to thank Joel Kastner and David Huenemoerder for useful discussion on the SER method,
 and the anonymous referee for useful comments.

Data presented herein were obtained at the W. M. Keck Observatory, which is operated as a scientific partnership among the 
California Institute of Technology, the University of California and the National Aeronautics and Space Administration. 
The Observatory was made possible by the generous financial support of the W.M. Keck Foundation. 
The authors wish to recognize and acknowledge the very significant cultural role and reverence that the summit of Mauna Kea has 
always had within the indigenous Hawaiian community.  
We are most fortunate to have the opportunity to conduct observations from this mountain. 
Part of this work has been supported by the National Science Foundation Science and Technology Center for Adaptive Optics, 
managed by the University of California at Santa Cruz under cooperative agreement AST 98-76783 and by the Packard Foundation.


\begin{table}
\caption{best fit to the zeroth order spectra of the HDE~245059 binary components}
\begin{center}
\begin{tabular}{ccc}
\tableline

Parameter  & north & south \\
\tableline

T$_1$ (MK)  &  $6.4^{+0.7}_{-0.5}$ &... \\
T$_2$ (MK)  &  $12.7 ^{+1.8}_{-0.9}$ & $8.3^{+0.6}_{-1.4}$ \\
T$_3$ (MK)  &  $39.7^{+27.3}_{-8.9}$ & $33.7 ^{+23.2} _{-11.6}$\\
T$_{\rm av}$ (MK) & 11.3 & 12.2\\

EM$_1$ ($ 10^{54}\rm{cm^{-3}}$) & $1.0^{+0.2}_{-0.1}$ & ... \\
EM$_2$ ($ 10^{54}\rm{cm^{-3}}$) & $0.9 \pm 0.2$ & $0.6 \pm 0.01$\\
EM$_3$ ($ 10^{54}\rm{cm^{-3}}$) & $0.4^{+0.1}_{-0.2}$ & $0.2 \pm 0.01$\\

$N_{\rm H}$ (10$^{19}$cm$^{-2}$) &:= 7.7 & :=7.7 \\
Flux ($10^{-13} \rm{ergs~cm^{-2}~s^{-1}}$) & 1.7  & 0.7\\
L$_{\rm X}$ ($10^{30}$ergs s$^{-1}$) & 3.3 & 1.4\\

C-statistics & 94 &  56\\
dof & 99 & 51\\

\tableline
  \end{tabular}
 \end{center}
 \tablecomments{
 We have included an absorption model with the value of the column density fixed to the best fit model to the quiescent state for the combined zeroth order plus grating spectra of the binary.
 }
 \label{table:ind_fits}
 \end{table}

\begin{table}
\caption{Best fit to the zeroth order spectrum of HDE~245059 south during the flare.}
\begin{center}
\begin{tabular}{cc}
\tableline
Parameter & value \\
\tableline
T (MK) & $30.4 ^{+12.6} _{-6.3}$\\
EM (10$^{54}$ cm$^{-3}$) & $1.3 ^{+0.3} _{-0.2}$ \\
$N_{\rm H}$ (10$^{19}$cm$^{-2}$) &  :=7.7 \\
Flux ($10^{-13} \rm{ergs~cm^{-2}~s^{-1}}$) & 3.9\\
L$_{\rm X}$ ($10^{30}$ erg s$^{-1}$) &  7.6 \\
 C-statistics & 8 \\
 dof &  11\\
\tableline
\end{tabular}
\end{center}
\tablecomments{The fit was obtained fixing all the parameters: temperatures, emission measures, abundances, and hydrogen column density to the values obtained from the best fit to the quiescent state and adding an extra temperature component which was set free to vary.}
\label{table:flare}
\end{table}

\begin{table}
\caption{Best fit 4-T model of the quiescent spectrum of the HDE~245059 binary using Method 1}
\begin{center}
\begin{tabular}{ccc}
\tableline
Parameter 		& Quiescent \\
\tableline
 T$_{1}$ (MK) 		& $3.9 \pm 0.3$\\
 T$_{2}$ (MK) 		& $8.1 \pm 0.3$\\
 T$_{3}$ (MK) 		& $15.6 ^{+1.2} _{-0.8}$ \\
 T$_{4}$ (MK) 		& $50.2 ^{+59} _{-8.1}$ \\
 T$_{\rm av}$ (MK) 	&  10.7\\
 
 EM$_{1}$ (10$^{54}$ cm$^{-3}$) 	& $1.1 \pm 0.2$  \\
 EM$_{2}$ (10$^{54}$ cm$^{-3}$) 	& $2.9 \pm 0.3$\\
 EM$_{3}$ (10$^{54}$ cm$^{-3}$) 	& $2.7 ^{+0.4} _{-0.3}$  \\
 EM$_{4}$ (10$^{54}$ cm$^{-3}$) 	& $0.6 ^{+0.2} _{-0.3}$ \\
 EM$_{\rm total}$ (10$^{54}$ cm$^{-3}$) & 7.29 \\
 $N_{\rm H}$ (10$^{19}$cm$^{-2}$) 	& $7.7^{+2.1} _{-2.6}$ \\
 
 O   	&  $0.30 _{-0.04} ^{+0.05}$\\
 Ne 	& $0.71 \pm 0.06 $ \\
 Mg 	& $0.21 \pm 0.03 $  \\
 Al  	& $0.34 _{-0.2} ^{+0.2} $\\
 Si  	& $0.18 \pm 0.02$ \\
 S   	& $0.14_{-0.04}  ^{+0.05}$ \\
 Ar  	& $0.46 \pm 0.3 $\\
 Ca  	&  $0.58 \pm 0.3$ \\
 Fe  	&  $0.23 \pm 0.02$ \\
 Ni  	& := Fe \\			  
 
 Flux ($10^{-13} \rm{ergs~cm^{-2}~s^{-1}}$) & 5.2\\
 L$_{\rm X}$ ($10^{31}$ erg s$^{-1}$) & 1.0\\
 
 C-statistics & 4148\\
 dof & 3708\\
\tableline
\end{tabular}
\end{center}
\tablecomments{Method 1 uses a discretization of the EMD by a multi-temperature model (see \ref{discrete} for details.)
We present here the best fit 4-T model of the average HDE~245059 binary for the quiescent state. 
We have used the combination of the zeroth order data plus the gratings HEG$\pm1$ and MEG$\pm1$. 
Abundances are given with respect to the solar photospheric values \citep{grev98}. All errors are calculated with $\Delta$C=1.}
\label{table:fits-total}
\end{table}


\begin{table}
\caption{Best fit to the quiescent spectrum of the HDE 245059 binary using Method 2}
\begin{center}
\begin{tabular}{cc}
\tableline
Parameter & value \\
\tableline
EM$_{\rm total}$ (10$^{54}$ cm$^{-3}$) & $7.3$\\  
$N_{\rm H}$ (10$^{19}$cm$^{-2}$) &  $6.8 ^{+2} _{-3}$  \\

O 	& $0.34 \pm 0.03  $ \\
Ne	& $0.69 \pm 0.06  $ \\
Mg	& $0.21 \pm 0.02  $ \\
Al	& $0.33 \pm 0.2$ \\
Si	& $0.18 \pm 0.02$ \\
S	& $0.15 ^{+0.05} _{-0.04}$ \\
Ar	& $0.46 \pm 0.3$ \\
Ca 	& $0.51 \pm 0.3$ \\
Fe	& $0.23 \pm 0.02$ \\
Ni	& = Fe \\

C-statistics & 4161 \\
dof &  3708\\
\tableline
\end{tabular}
\end{center}
\tablecomments{Method 2 is based on a continuous EMD obtained from Chebychev polynomials, in this case we have used a degree of 8 (see \ref{cheby}).
This method has been applied to the quiescent state spectrum of the binary. Abundances are given with respect to the solar photospheric values \citep{grev98}. Errors are calculated with $\Delta$C=1.}
\label{table:chebychev}
\end{table}


\begin{table}
\caption{Best fit to the quiescent spectrum of the HDE~245059 binary using Method 3 }
\begin{center}
\begin{tabular}{ccc}
\tableline
Parameter  &$\alpha$ fixed & $\alpha$, $\beta$ free \\
\tableline
$\alpha$   & :=2.0 & $2.4 \pm 0.4$\\
$\beta$	& $-1.5 \pm 0.09$ & $-1.5 \pm 0.09$\\
${\rm log}\, {\rm T}_0$	& $7.0 \pm 0.01$ &	$6.9 ^{+0.03} _{-0.02}$\\
EM$_{\rm total}$ (10$^{54}$ cm$^{-3}$) &  7.49 & 7.31 \\
$N_{\rm H}$ (10$^{19}$cm$^{-2}$) &  $8.7 \pm 2.5$  & $8.0 ^{+2.6} _{-2.5}$\\

O 	& $0.28 \pm 0.04  $ & $0.30 ^{+0.05} _{-0.04}$\\
Ne	& $0.68 ^{+0.06} _{-0.05}$ & $0.69 ^{+0.06} _{-0.05}$\\
Mg	& $0.21 ^{+0.03} _{-0.02}$ & $0.20 ^{+0.03} _{0.02}$\\
Al	& $0.35 \pm 0.2$   & $0.34 \pm 0.2$\\
Si	& $0.18 \pm 0.02$ & $0.17 \pm 0.02$\\
S	& $0.15 ^{+0.05} _{-0.04}$  & $0.15 ^{+0.05} _{-0.04}$\\
Ar	& $0.43 \pm 0.3$ & $0.44 \pm 0.3$ \\
Ca 	& $0.45 \pm 0.3$ & $0.46 \pm 0.3$\\
Fe	& $0.23 \pm 0.02$ & $0.22 \pm 0.02$\\
Ni	& = Fe & = Fe\\

C-statistics & 4168 & 4167 \\
dof &  3714 & 3713\\
\tableline
\end{tabular}
\end{center}
\tablecomments{Method 3 is based on a continuous EMD described by two power laws. 
It has been applied to the quiescent spectrum of the binary. In the first case we have fixed one of them, while in the second case we have left both the power law indexes free to vary (see \ref{powerlaw} for details). Abundances are given with respect to the solar photospheric values \citep{grev98}
Errors are calculated with $\Delta$C=1.}
\label{table:PL}
\end{table}

\begin{table}
\caption{Line fluxes for the binary HDE~245059 }
\begin{center}
\begin{tabular}{ccc}
\tableline
\tableline
Ion & $\lambda$ (\AA) & Flux ($10^{-5}$ photons cm$^{-2}$ s$^{-1}$) \\
\tableline
Si~XIII(r) & $6.65$ & $0.5 \pm 0.1$ \\
Si~XIII(i) & $6.68$ & $0.1 \pm 0.04$\\
Si~XIII(f) & $6.74$ & $0.3 \pm 0.1$ \\
\tableline
Mg~XI(r)   & $9.17$ & $0.3 \pm 0.1$\\
Mg~XI(i)   & $9.23$ & $0.2 \pm 0.1$\\
Mg~XI(f)   & $9.31$ & $0.3 \pm 0.1$\\
\tableline 
Ne~IX(r)   & $13.45$ & $2.7 \pm 0.3$\\
Ne~IX(i)   & $13.55$ & $0.9 \pm 0.2$\\
Ne~IX(f)   & $13.70$ & $1.7 \pm 0.3$\\
Fe~XIX     & $13.52$ & $1.4 \pm 0.3$ \\
\tableline
 &  & 0.68 confidence \\
O~VII(r) & 21.6 & $<1.4$ \\
O~VII(i) & 21.8 & $<1.2$ \\
O~VII(f) & 22.1 & $<2.1$ \\

\tableline
& & 0.90 confidence \\
O~VII(r) & 21.6 & $<2.3$ \\
O~VII(i) & 21.8 & $<2.1$ \\
O~VII(f) & 22.1 & $<3.5$ \\
 
\tableline

Ne~X~Ly$\alpha$ & $12.13$ & $10.3 _{-0.5} ^{+0.4}$ \\
Fe~XVII    & $15.01$ & $6.7 _{-0.6} ^{+0.5}$ \\
O~VIII~Ly$\alpha$ & $18.97$ & $14.2 _{-1.8} ^{+1.4}$\\
\tableline
\end{tabular}
\end{center}
\tablecomments{Single line fluxes and He-like triplets obtained from the model of the MEG$\pm1$ and HEG$\pm1$ data using the combination of a bremsstrahlung model for the continuum and delta profiles for the lines. 
For the O~VII triplet we include the flux estimate at 2 confidence levels; 68\% and 90\%. 
In the case of Ne IX we have also added a delta profile for Fe XIX which is blended with the Ne line. 
The fluxes are not corrected for the absorption column density.} 
\label{table:helikebinary}
\end{table}


\begin{table}
\caption{Line flux ratios and derived electron densities for the HDE~245059 binary}
\begin{center}
\begin{tabular}{ccccc}
\tableline
           & \multicolumn{2}{c}{0.68 confidence} &\multicolumn{2}{c}{0.90 confidence} \\
 \tableline
 Triplet & R & $n_{e}$ (cm$^{-3}$) &R & $n_{e}$ (cm$^{-3}$) \\
 \tableline
 Ne IX & $2.0 \pm 0.8$ & $5 \pm 5 \times 10^{11}$ & $2.0 \pm 1.1$ & $< 2 \times 10^{12}$ \\
 Mg XI &$1.4 \pm 0.6$ & $1^{+3} _{-0.5} \times 10^{13}$ & $1.4 \pm 1.0$ & $1 ^{+6} _{-0.8} \times 10^{13}$ \\
 Si XIII  &$2.3 \pm 0.9$ & $< 5 \times 10^{13}$ & $2.3 \pm 1.6$ & $< 3 \times 10^{14}$ \\
\hline
\end{tabular}
\end{center}
\tablecomments{Line flux ratios $R=f/i$ and electron densities derived from the He-like triplets. 
Calculations were made for the grating spectrum of the binary  (HEG and MEG first order) at two confidence levels, 68\% and \%90. 
See section \ref{density} for discussion.
} 
\label{table:densities}
\end{table}

\begin{table}
\caption{Line fluxes for HDE~245059 binary components}
\begin{center}
\begin{tabular}{cccc}
\tableline
& &  \multicolumn{2}{c}{Flux ($10^{-5}$photons~cm$^{-2}$~s$^{-1}$)} \\
\multicolumn{1}{c}{Ion} &\multicolumn{1}{c}{$\lambda$ (\AA)} &  \multicolumn{1}{c}{north} & \multicolumn{1}{c}{south} \\
\tableline
Mg~XII           & 8.42  & $0.5 \pm 0.1$ & $0.1 \pm 0.1$  \\
Ne~X~Ly$\alpha$ & 12.13 & $3.7 \pm 0.4$ & $2.8 \pm 0.4$ \\ 
Ne~X~Ly$\beta$  & 10.24 & $0.5 \pm 0.1$ & $0.2 \pm 0.1$  \\
O~VIII~Ly$\alpha$& 18.97 & $7.4 ^{+2} _{-1}$ & $5.7 ^{+2} _{-1}$ \\
\hline
\end{tabular}
\end{center}
\tablecomments{Single line flux obtained for each star of the binary from the model of the MEG$\pm1$ data using the combination of a bremsstrahlung model for the continuum 
and a delta function for the line profiles. 
The fluxes are not corrected for the absorption column density. 
} 
\label{table:NS_flux}
\end{table}


\end{document}